\documentstyle[preprint,pre,aps,eqsecnum]{revtex}
\begin{document}
\draft
\title{Period $p$-tuplings in coupled maps}
\author{Sang-Yoon Kim
\footnote{Permanent address: Department of Physics, Kangwon National
          University, Chunchon, Kangwon-Do 200-701, Korea.
          Electronic address: sykim@cc.kangwon.ac.kr}
       }
\address{School of Physics, Georgia Institute of Technology,
         Atlanta, GA 30332-0430}
\maketitle

\begin{abstract}
We study the critical behavior (CB) of all period $p$-tuplings
$(p \!=\!2,3,4,\dots)$ in $N$ $(N \!=\! 2,3,4,\dots)$ symmetrically
coupled one-dimensional maps.
We first investigate the CB for the $N=2$ case of two coupled maps,
using a renormalization method. Three (five) kinds of fixed points of
the renormalization transformation and their relevant
``coupling eigenvalues'' associated with coupling perturbations are
found in the case of even (odd) $p$. We next study the CB for the
linear- and nonlinear-coupling cases (a coupling is called linear or
nonlinear according to its leading term), and confirm the
renormalization results. Both the structure of the critical set (set
of the critical points) and the CB vary according as the coupling is
linear or nonlinear. Finally, the results of the two coupled maps are
extended to many coupled maps with $N \geq 3$, in which  the CB
depends on the range of coupling.
\end{abstract}


%
% End of Abstract
%

\narrowtext

\section{Introduction}

\label{sec:Int}

Universal scaling behavior of period $p$-tuplings $(p=2,3,4,\dots)$
has been found in a one-parameter family $f_A(x)$ of one-dimensional
(1D) unimodal maps with a quadratic maximum. As the nonlinearity
parameter $A$ increases, a stable fixed point undergoes
the cascade of period-doubling bifurcations accumulating at a finite
parameter value $A_{\infty}$. The period-doubling sequence
corresponding to the MSS (Metropolis, Stein, and Stein
\cite{Metropolis}) sequence $R^{*n}$ (for details of the MSS sequences
and the $(*)$-composition rule, see Refs.~\cite{Metropolis,Derrida})
exhibits an asymptotic scaling behavior \cite{Feigenbaum}.

What happens beyond the period-doubling accumulation point $A_\infty$
is interesting from the viewpoint of chaos. The parameter interval
between $A_{\infty}$ and the final boundary-crisis point $A_c$ beyond
which no periodic or chaotic attractors can be found within the
unimodality interval is called the ``chaotic'' regime. Within this
region, besides the period-doubling sequence, there are many other
sequences of periodic orbits exhibiting their own scaling behavior.
In particular, every primary pattern $P$ (that cannot be decomposed
using the ($*$)-operation) leads to an MSS sequence $P^{*n}$. For
example, $P=RL$ leads to a period-tripling sequence,
$P=RL^2$ to a period-quadrupling sequence, and the three different
$P= RLR^2,\;RL^2R$, and $RL^3$ to three different period
$5$-tupling sequences. Thus there exist infinitely many higher period
$p$-tupling $(p=3,4,\dots)$ sequences inside the chaotic regime.
Unlike the period-doubling sequence, stability regions of periodic
orbits in the higher period $p$-tupling sequences are not adjacent on
the parameter axis, because they are born by their own tangent
bifurcations. The asymptotic scaling behavior of these (disconnected)
higher period $p$-tupling sequences characterized by the
parameter and orbital scaling factors, $\delta$ and $\alpha$, vary
depending on the primary pattern $P$
\cite{Derrida,Hu,Eckmann,Hao,Hauser,Chang,Kenny,Urumov}.

In this paper we consider $N$ $(N=2,3,4,\dots)$ symmetrically coupled
1D maps, which may be used as models of coupled nonlinear oscillators
such as Josephson-junction arrays or chemically reacting cells, and so
on \cite{Kaneko}. We are interested in the critical behavior (CB) of
period $p$-tuplings $(p=2,3,\dots)$ in the coupled 1D maps. The
period-doubling case with $p=2$ was previously studied \cite{KK,Kim1}.
Here we study the critical scaling behavior of all the other higher
period $p$-tuplings ($p=3,4,\dots$).

Using a renormalization method, we first investigate the critical
behavior for the $N=2$ case of two coupled maps in
Sec.~\ref{sec:RATCM}. In the case of even (odd) $p$, we find three
(five) kinds of fixed points of a renormalization transformation and
their relevant coupling eigenvalues (CE's) associated with coupling
perturbations. A short account of the renormalization result has been
already published \cite{Kim2}. We next consider two kinds of
couplings, linear- and nonlinear-coupling cases; a coupling is called
linear or nonlinear according to its leading term. As examples of the
linear- and nonlinear-coupling cases, we study the linearly- and
dissipatively-coupled maps, respectively in Sec.~\ref{sec:EX}, and
confirm the renormalization results. The structure of the critical set
(set of the critical points) varies depending on the nature of
coupling. In the linearly-coupled case, an infinite number of critical
line segments and the zero-coupling critical point, at which the $N$
1D maps become uncoupled, constitute the critical set, while in the
dissipatively-coupled case, the critical set consists of
only one critical line segment, one end of which is the zero-coupling
critical point. The CB also depends on the position on
the critical set. For even (odd) $p$, three (four) kinds of fixed
points govern the CB for the linearly-coupled case, whereas only two
(three) fixed points govern the CB for the dissipatively-coupled case.

In Sec.~\ref{sec:EXMCM} we extend the results of the two coupled maps
to many coupled maps with $N \geq 3$. It is found that the critical
scaling behavior depends on the range of coupling. In the extreme
long-range case of global coupling, in which each 1D map is
coupled to all the other 1D maps with equal strength, both the
structure of the critical set and the CB are the same
as those for the two-coupled case, independently of $N$.
However, for the cases of nonglobal couplings of shorter range,
a significant change in the structure of the critical set may
or may not occur according as the coupling is linear or not.
For the case of a linear nonglobal coupling, only the
zero-coupling critical point is left in the parameter plane, which is
in contrast to the global-coupling case. On the other hand, for the
case of a nonlinear nonglobal coupling, one critical line segment
still remains, as in the globally-coupled case.
Finally, a summary is given in Sec.~\ref{sec:SUM}.

\section{Renormalization analysis of two coupled maps}
\label{sec:RATCM}

In this section we first discuss stability of periodic orbits in
two coupled 1D maps, and then study the CB associated
with period $p$-tuplings $(p=2,3,4,\dots)$ using a renormalization
method. Three (five) kinds of fixed points of a renormalization
operator and their relevant CE's with moduli larger than unity are
found in the case of even (odd) $p$.

\subsection{Stability of periodic orbits in two coupled maps}
\label{sub:TCM}

We consider a map $T$ consisting of two symmetrically coupled 1D maps,
\begin{equation}
T:\left\{ 
\begin{array}{l}
x_{t+1} = F(x_t,y_t) = f(x_t) + g(x_t,y_t), \\ 
y_{t+1} = F(y_t,x_t) = f(y_t) + g(y_t,x_t),
\end{array}
\right.
\label{eq:CM}
\end{equation}
where $t$ denotes a discrete time, $f(x)$ is a 1D unimodal map with a
quadratic maximum at $x=0$, and $g(x,y)$ is a coupling function. The
uncoupled 1D map $f$ satisfies a normalization condition
\begin{equation}
f(0) = 1,  \label{eq:NC}
\end{equation}
and the coupling function $g$ obeys a condition 
\begin{equation}
g(x,x) = 0 \;\;\;{\rm for \; any \;} x.  \label{eq:CC}
\end{equation}

The two-coupled map (\ref{eq:CM}) is invariant under the exchange of
coordinates such that $x \leftrightarrow y$. The set of all points
which are invariant under the exchange of coordinates forms a symmetry
line $y=x$. An orbit is called a(n) (in-phase) synchronous orbit if it
lies on the symmetry line, i.e., it satisfies
\begin{equation}
x_t = y_t\;\;\;{\rm for \; all\;} t.  \label{eq:IO}
\end{equation}
Otherwise, it is called an (out-of-phase) asynchronous orbit. Here we
study only the synchronous orbits, which can be easily found from the
uncoupled 1D map, $x_{t+1} = f(x_t)$, because of the condition
(\ref{eq:CC}).

Stability analysis of a periodic orbit in the two-coupled map $T$ can
be conveniently carried out in a set of new coordinates $(X,Y)$,
defined by
\begin{equation}
  X = {(x+y) \over 2}, \;\; Y = { (x-y) \over 2}.
\label{eq:XY}
\end{equation}
Here $X$ and $Y$ correspond to the synchronous
and asynchronous modes of the orbit, respectively.
For example, for a synchronous orbit $X=x$ and $Y=0$, whereas for
an asynchronous orbit $Y \neq 0$.

In order to study the stability of a synchronous orbit with period
$q$, we consider an infinitesimal perturbation $(\delta X, \delta Y)$
to the orbit. Here $\delta X$ and $\delta Y$ correspond to the
synchronous-mode and asynchronous-mode perturbations, respectively.
Linearizing the $q$-times iterated map $T^q$
(expressed in terms of the new coordinates) at an orbit point,
we obtain a linearized map,
\begin{equation}
 \left( \begin{array}{c}
         \delta X_{t+q}  \\
         \delta Y_{t+q}
      \end{array}
      \right)
      = J
      \left( \begin{array}{c}
         \delta X_t  \\
         \delta Y_t
      \end{array}
      \right),
\label{eq:LM}
\end{equation}
where the Jacobian matrix $J$ $(\equiv DT^q)$ of $T^q$ is given by the
$q$-product of the Jacobian matrix $DT$ of $T$ along the orbit:
\begin{eqnarray}
J &=& \prod_{t=0}^{q-1} DT(x_t,x_t) \nonumber \\
 &=&  \prod_{t=0}^{q-1}
   \left(
   \begin{array} {cc}
   f'(x_t) &  0 \\
   0 & f'(x_t) - 2 G(x_t)
   \end{array}
   \right).
\label{eq:J}
\end{eqnarray}
Here the prime denotes a derivative with respect to $x$, and
$G(x) = {\partial g}(x,y)/{\partial y} \mid_{y=x}$; hereafter,
$G(x)$ will be referred to as the ``reduced coupling function'' of
$g(x,y)$. Note that $\delta X$ and $\delta Y$ become decoupled for
the case of a synchronous orbit (i.e., $J$ has a diagonalized form).

The eigenvalues of $J$, called the stability multipliers of
the orbit, are then given by:
\begin{equation}
\lambda_0 = \prod_{t=0}^{q-1} f^{\prime}(x_t), \;\;\; \lambda_1 =
\prod_{t=0}^{q-1} [f^{\prime}(x_t)-2G(x_t)].
\label{eq:MULTI}
\end{equation}
The two stability multipliers, $\lambda_0$ and
$\lambda_1$, determine the stability of the synchronous orbit
against the synchronous-mode and asynchronos-mode perturbations,
respectively. Hereafter, they will be called the synchronous and
asynchronous stability multipliers, respectively.
Note also that the synchronous stability multiplier $\lambda_0$ is
just the stability multiplier of the uncoupled 1D map, and the
coupling affects only the asynchronous stability multiplier
$\lambda_1$.

A synchronous orbit is stable when it is stable against both the
synchronous-mode and asynchronous-mode perturbations, i.e.,
the moduli of both stability multipliers are less than unity
($|\lambda_i| < 1$ for $i=0,1$). Hence, the stable region of a
synchronous orbit in the parameter plane is bounded by the
synchronous and asynchronous bifurcation lines determined by the
equations $\lambda_i= \pm 1$ for $i=0,1$, as will be seen in
Sec.~\ref{sec:EX}. When the $\lambda_0=1$ $(-1)$
line is crossed, the synchronous orbit loses its stability via
synchronous saddle-node (period-doubling) bifurcation.
On the other hand, when the $\lambda_1=1$ $(-1)$ line is crossed,
it becomes unstable via asynchronous pitchfork (period-doubling)
bifurcation. Some brief explanations on the bifurcations are
given below.

In the case of a synchronous saddle-node bifurcation, the synchronous
orbit collides with an unstable synchronous orbit with the same
period, and then they disappear, like the tangent bifurcation in the
1D maps. On the other hand, there are two types of supercritical and
subcritical bifurcations for each case of the pitchfork and
period-doubling bifurcations. In the supercritical case of the
synchronous (asynchronous) pitchfork and period-doubling bifurcations,
the synchronous orbit loses its stability, and gives rise
to the birth of a pair of new stable synchronous (asynchronous) orbits
with the same period and a new stable synchronous (asynchronous)
period-doubled orbit, respectively. However, in the subcritical case
of the synchronous (asynchronous) pitchfork and period-doubling
bifurcations, the synchronous orbit becomes unstable by
absorbing a pair of unstable synchronous (asynchronous) orbits with
the same period and an unstable synchronous (asynchronous)
period-doubled orbit, respectively. (For more details on
bifurcations on 2D dissipative maps, refer to \cite{Gukenheimer}.)

\subsection{Renormalization analysis}
\label{sub:RA}

We now consider the period $p$-tupling renormalization transformation
${\cal N}$, which is composed of the $p$-times iterating $(T^{(p)})$
and rescaling $(B)$ operators:
\begin{equation}
{\cal N}(T) \equiv B T^{(p)} B^{-1}.  \label{eq:RON}
\end{equation}
Here the rescaling operator $B$ is: 
\begin{equation}
B = \left( 
\begin{array}{cc}
\alpha & \;\;\; 0 \\ 
0 & \;\;\; \alpha
\end{array}
\right),
\label{eq:SO}
\end{equation}
because we consider only synchronous orbits.

Applying the renormalization operator ${\cal N}$ to the coupled map
(\ref {eq:CM}) $n$ times, we obtain the $n$-times renormalized map
$T_n$ of the form,
\begin{equation}
{T_n}:\left\{ 
\begin{array}{l}
x_{t+1} = {F_n}(x_t,y_t) = {f_n}(x_t) + {g_n}(x_t,y_t), \\ 
y_{t+1} = {F_n}(y_t,x_t) = {f_n}(y_t) + {g_n}(y_t,x_t).
\end{array}
\right.  \label{eq:RTn}
\end{equation}
Here ${f_n}$ and ${g_n}$ are the uncoupled and coupling parts of the
$n$-times renormalized function $F_n$, respectively. They satisfy the
following recurrence equations:
\begin{eqnarray}
&f_{n+1}(x) = & \alpha f_n^{(p)} ({\frac {x} {\alpha}}),
\label{eq:RUCFn} \\
&g_{n+1}(x,y) =& \alpha F_n^{(p)} ({\frac {x} {\alpha}}, {\frac {y}
{\alpha}}) -{\alpha} f_n^{(p)} ({\frac {x} {\alpha}}), \label{eq:RCFn}
\end{eqnarray}
where
\begin{eqnarray}
&f_n^{(p)} (x)& = f_n (f_n^{(p-1)}(x)), \\
&F_n^{(p)}(x,y)& = F_n(F_n^{(p-1)}(x,y), F_n^{(p-1)}(y,x)),
\end{eqnarray}
and the rescaling factor is chosen
to preserve the normalization condition $f_{n+1}(0)=1$, i.e.,
\begin{equation}
\alpha = {1 \over f^{(p-1)}_n(1)}.
\end{equation}
The recurrence relations (\ref{eq:RUCFn}) and (\ref{eq:RCFn}) define a
renormalization operator ${\cal R}$ of transforming a pair of
functions $(f,g)$:
\begin{equation}
\left( 
\begin{array}{c}
{f_{n+1}} \\ 
{g_{n+1}}
\end{array}
\right) = {\cal R} \left( 
\begin{array}{c}
{f_n} \\ 
{g_n}
\end{array}
\right).  \label{eq:RORn}
\end{equation}

A map $T_c$ with the nonlinearity and coupling parameters set to their
critical values is called a critical map:
\begin{equation}
{T_c}:\left\{
\begin{array}{l}
x_{t+1} = {F_c}(x_t,y_t) = {f_c}(x_t) + {g_c}(x_t,y_t), \\
y_{t+1} = {F_c}(y_t,x_t) = {f_c}(y_t) + {g_c}(y_t,x_t).
\end{array}
\right.  \label{eq:CRM}
\end{equation}
A critical map is attracted to a fixed map $T^*$ under iterations of
the renormalization transformation $\cal N$:
\begin{equation}
{T^*}:\left\{
\begin{array}{l}
x_{t+1} = {F^*}(x_t,y_t) = {f^*}(x_t) + {g^*}(x_t,y_t), \\
y_{t+1} = {F^*}(y_t,x_t) = {f^*}(y_t) + {g^*}(y_t,x_t).
\end{array}
\right.  \label{eq:FM}
\end{equation}
Here $(f^*,g^*)$ is a fixed point of the renormalization operator
$\cal R$ with $\alpha = 1/ f^{*(p-1)}(1)$:
\begin{equation}
\left( 
\begin{array}{c}
{f^*} \\
{g^*}
\end{array}
\right) = {\cal R} \left( 
\begin{array}{c}
{f^*} \\
{g^*}
\end{array}
\right).  \label{eq:FPEQ}
\end{equation}
This fixed-point equation can be solved row by row consecutively.
Note that $f^*(x)$ is just the fixed function in the 1D map case,
which varies depending on $p$ \cite{Eckmann,Hao,Chang,Kenny}. Only
the equation for the coupling fixed function $g^*(x,y)$ is therefore
left to be solved. One trivial solution is $g^*(x,y)=0$. In this
zero-coupling case, the fixed map (\ref{eq:FM}) consists of two
uncoupled 1D fixed maps, which is associated with the CB at the
zero-coupling critical point.

However, it is not easy to directly find coupling fixed functions
other than the zero-coupling fixed function $g^*(x,y)=0$.
We therefore introduce a tractable recurrence equation for a reduced
coupling function $G(x) = \partial g(x,y) / \partial y |_{y=x}$.
Differentiating the recurrence equation (\ref{eq:RCFn}) for
$g(x,y)$ with respect to $y$ and setting $y=x$, we obtain a
recurrence equation for $G(x)$ \cite{note}:
\begin{eqnarray}
G_{n+1}(x) &=& F_{n,2}^{(p)}({x \over \alpha}) \nonumber \\
&=& F_{n,2}^{(p-1)} ({x \over \alpha})
[{f'_n}({f_n^{(p-1)}}({x \over \alpha}))-
2G_{n}({f_n^{(p-1)}}({x \over \alpha}))] \nonumber \\
&& +
f^{(p-1)'}_n({x \over \alpha}) G_{n}({f_n^{(p-1)}}({x \over \alpha})),
\label{eq:RRE}
\end{eqnarray}
where  $F^{(p)}_{n,2}(x) \equiv \partial F^{(p)}_n (x,y) /
\partial y|_{y=x}$.
Then Eqs.~(\ref{eq:RUCFn}) and (\ref{eq:RRE})
define a ``reduced renormalization operator'' $\tilde{\cal R}$
of transforming a pair of functions
$(f,G)$:
\begin{equation}
\left( 
\begin{array}{c}
{f_{n+1}} \\ 
{G_{n+1}}
\end{array}
\right) = {\tilde{\cal R}} \left(
\begin{array}{c}
{f_n} \\ 
{G_n}
\end{array}
\right).  \label{eq:RRORn}
\end{equation}
We look for fixed points $({f^*},{G^*})$ of ${\tilde {\cal R}}$, which
satisfy
\begin{equation}
\left( 
\begin{array}{c}
{f^*} \\
{G^*}
\end{array}
\right) = {\tilde{\cal R}} \left(
\begin{array}{c}
{f^*} \\
{G^*}
\end{array}
\right).  \label{eq:RFPEQ}
\end{equation}
Here $f^*$ is just the 1D fixed function and $G^*$ is the reduced
coupling fixed function of $g^*$, i.e.,
${G^*}(x) = {\partial {g^*}(x,y)}/{\partial y} \mid_{y=x}$.

We first consider the lowest two cases with $p=2,3$.
For $p=2$, Eq.~(\ref{eq:RRE}) becomes
\begin{eqnarray}
G_{n+1}(x) &=& F_{n,2}^{(2)}({x \over \alpha}) \nonumber \\
&=& G_n({x \over \alpha})
[{f'_n}(f_n({x \over \alpha}))-
2G_{n}({f_n}({x \over \alpha}))] \nonumber \\
&& +
f'_n({x \over \alpha}) G_{n}({f_n}({x \over \alpha})).
\label{eq:p2RRE}
\end{eqnarray}
We note that $F_{n,2}^{(2)} ({x \over \alpha})$ in
Eq.~(\ref{eq:p2RRE}) becomes
\begin{mathletters}
\label{eq:p2F2}
\begin{eqnarray}
F_{n,2}^{(2)} ({x \over \alpha}) &=& 0 ~~~{\rm for}~G_n(x)=0, \\
F_{n,2}^{(2)} ({x \over \alpha}) &=& {1 \over 2} f_n^{(2)'}
({x \over \alpha})~~~{\rm for}~G_n(x)= {1 \over 2} f'_n(x), \\
F_{n,2}^{(2)} ({x \over \alpha}) &=& {1 \over 2}
                        [f_n^{(2)'}({x \over \alpha})-1] \nonumber \\
&&     ~~~{\rm for}~G_n(x)= {1 \over 2} [f'_n(x)-1], \\
F_{n,2}^{(2)} ({x \over \alpha}) &=& {1 \over 2}
                        [f_n^{(2)'}({x \over \alpha})-1] \nonumber \\
 &&       ~~~{\rm for}~G_n(x)= {1 \over 2} [f'_n(x)+1], \\
F_{n,2}^{(2)} ({x \over \alpha}) &=& 0
              ~~~{\rm for}~G_n(x)= f'_n(x).
\end{eqnarray}
\end{mathletters}
\noindent
Differentiating the 1D fixed-function equation for $f^*(x)$ with
respect to $x$ in the period $p$-tupling case, we have
\begin{equation}
f^{*'}(x) = f^{*(p)'}({x \over \alpha})=
\prod_{i=0}^{p-1} f^{*'}(f^{*(i)}({x \over \alpha})).
\label{eq:1DD}
\end{equation}
Using Eqs.~(\ref{eq:p2F2}) and (\ref{eq:1DD}), we obtain three
solutions for the reduced coupling fixed function $G^*(x)$:
\begin{mathletters}
\begin{eqnarray}
G^*(x) &=& 0, \\
G^*(x) &=& {1 \over 2} f^{*'}(x), \\
G^*(x) &=& {1 \over 2} [f^{*'}(x) - 1].
\end{eqnarray}
\end{mathletters}
Thus we find three fixed points $(f^*,G^*)$ of $\tilde {\cal R}$ for
the period-doubling case.

We next consider the period-tripling case with $p=3$.
The recurrence equation for $G$ becomes
\begin{eqnarray}
G_{n+1}(x) &=& F_{n,2}^{(3)}({x \over \alpha}) \nonumber \\
&=& F_{n,2}^{(2)}({x \over \alpha})
[{f'_n}(f_n^{(2)}({x \over \alpha}))-
2G_{n}({f_n}^{(2)}({x \over \alpha}))] \nonumber \\
&& +
f^{(2)'}_n({x \over \alpha}) G_{n}({f_n^{(2)}}({x \over \alpha})).
\label{eq:p3RRE}
\end{eqnarray}
We also note that $F_{n,2}^{(3)} ({x \over \alpha})$ in
Eq.~(\ref{eq:p3RRE}) becomes [see Eq.~(\ref{eq:p2F2}) for
$F_{n,2}^{(2)}$]
\begin{mathletters}
\label{eq:p3F2}
\begin{eqnarray}
F_{n,2}^{(3)} ({x \over \alpha}) &=& 0 ~~~{\rm for}~G_n(x)=0, \\
F_{n,2}^{(3)} ({x \over \alpha}) &=& {1 \over 2} f_n^{(3)'}
({x \over \alpha})
              ~~~{\rm for}~G_n(x)= {1 \over 2} f'_n(x), \\
F_{n,2}^{(3)} ({x \over \alpha}) &=& {1 \over 2}
                        [f_n^{(3)'}({x \over \alpha})-1] \nonumber \\
&&    ~~~{\rm for}~G_n(x)= {1 \over 2} [f'_n(x)-1], \\
F_{n,2}^{(3)} ({x \over \alpha}) &=& {1 \over 2}
                        [f_n^{(3)'}({x \over \alpha})+1] \nonumber \\
&&    ~~~{\rm for}~G_n(x)= {1 \over 2} [f'_n(x)+1], \\
F_{n,2}^{(3)} ({x \over \alpha}) &=& f_n^{(3)'}({x \over \alpha})
              ~~~{\rm for}~G_n(x)= f'_n(x).
\end{eqnarray}
\end{mathletters}
\noindent
Using Eqs.~(\ref{eq:1DD}) and (\ref{eq:p3F2}), we get five solutions
for $G^*(x)$:
\begin{mathletters}
\begin{eqnarray}
G^*(x) &=& 0, \\
G^*(x) &=& {1 \over 2} f^{*'}(x), \\
G^*(x) &=& {1 \over 2} [f^{*'}(x) - 1], \\
G^*(x) &=& {1 \over 2} [f^{*'}(x) + 1], \\
G^*(x) &=& f^{*'}(x).
\end{eqnarray}
\end{mathletters}
\noindent
Thus we find five kinds of fixed points $(f^*,G^*)$ for the
period-tripling case.

We now consider the general period p-tupling case. By induction, we
obtain the solutions for $G^*(x)$ as follows. Assume that
$F_{n,2}^{(p)}({x \over \alpha})$ for the case of period $p$-tupling
becomes
\begin{mathletters}
\label{eq:pnF2}
\begin{eqnarray}
F_{n,2}^{(p)} ({x \over \alpha}) &=& 0 ~~~{\rm for}~G_n(x)=0, \\
F_{n,2}^{(p)} ({x \over \alpha}) &=& {1 \over 2} f_n^{(p)'}({x \over
\alpha})~~~{\rm for}~G_n(x)= {1 \over 2} f'_n(x), \\
F_{n,2}^{(p)} ({x \over \alpha}) &=& {1 \over 2}
                        [f_n^{(p)'}({x \over \alpha})-1] \nonumber \\
&&        ~~~{\rm for}~G_n(x)= {1 \over 2} [f'_n(x)-1], \\
F_{n,2}^{(p)} ({x \over \alpha}) &=&
 \left \{
 \begin{array}{cl}
 {1 \over 2} [f_n^{(p)'}({x \over \alpha})-1] ~&{\rm for~even~} p \\
 {1 \over 2} [f_n^{(p)'}({x \over \alpha})+1] ~&{\rm for~odd~} p
 \end{array}
 \right. \nonumber \\
&&~~~{\rm for}~G_n(x)= {1 \over 2} [f'_n(x)+1], \\
F_{n,2}^{(p)} ({x \over \alpha}) &=&
  \left \{
  \begin{array}{cl}
   0~&{\rm for~even~} p \\
   f_n^{(p)'}({x \over \alpha})~&{\rm for~odd~} p
  \end{array}
\right. \nonumber \\
&& ~~~{\rm for}~ G_n(x)=f_n'(x).
\end{eqnarray}
\end{mathletters}
\noindent
For the case of the next period $(p+1)$-tupling, the recurrence
equation (\ref{eq:RRE}) for $G$ becomes
\begin{eqnarray}
G_{n+1}(x) &=& F_{n,2}^{(p+1)}({x \over \alpha}) \nonumber \\
&=& F_{n,2}^{(p)}({x \over \alpha})
[{f'_n}(f_n^{(p)}({x \over \alpha}))-
2G_{n}({f_n}^{(p)}({x \over \alpha}))] \nonumber \\
&& +
f^{(p)'}_n({x \over \alpha}) G_{n}({f_n^{(p)}}({x \over \alpha})).
\label{eq:ppRRE}
\end{eqnarray}
It follows from the assumption (\ref{eq:pnF2}) that
$F_{n,2}^{(p+1)}({x \over \alpha})$ for
the case of period $(p+1)$-tupling also satisfies Eq.~(\ref{eq:pnF2}).
Note that the lowest even and odd cases, $p=2$ and $3$, satisfy the
assumption (\ref{eq:pnF2}) [see Eqs.~(\ref{eq:p2F2}) and
(\ref{eq:p3F2})].
Consequently, the equation (\ref{eq:pnF2}) is valid for all $p$.
Using Eqs.~(\ref{eq:1DD}) and (\ref{eq:pnF2}), we obtain three (five)
solutions for $G^*(x)$ in the case of even (odd) $p$:
\begin{mathletters}
\label{eq:RCFFTN}
\begin{eqnarray}
G^*(x) &=& 0~~~{\rm for~all~}p, \\
G^*(x) &=& {1 \over 2} f^{*'}(x)~~~{\rm for~all~}p, \\
G^*(x) &=& {1 \over 2} [f^{*'}(x) - 1]~~~{\rm for~all~}p, \\
G^*(x) &=& {1 \over 2} [f^{*'}(x) + 1]~~~{\rm for~odd~}p, \\
G^*(x) &=& f^{*'}(x) ~~~{\rm for~odd~}p.
\end{eqnarray}
\end{mathletters}
\noindent
Thus we find three (five) kinds of fixed points $(f^*,G^*)$ of
$\tilde {\cal R}$ for the case of even (odd) $p$.

In case of a critical map (\ref{eq:CRM}), the synchronous and
asynchronous stability multipliers $\lambda_{0,n}$ and $\lambda_{1,n}$
of the synchronous orbits are
given by [see Eq.~(\ref{eq:MULTI})]
\begin{equation}
\lambda_{0,n} = \prod_{t=0}^{q-1} f'_c(x_t), \;\;\;
\lambda_{1,n} = \prod_{t=0}^{q-1}
[f'_c(x_t)-2G_c(x_t)],
\end{equation}
where $G_c(x)$ is the reduced coupling function of $g_c(x)$, i.e.,
$G_c(x)= {\partial g_c} / {\partial y} |_{y=x}$.
As $n \rightarrow \infty$, they converge to their limit values
$\lambda_0^*$ and $\lambda_1^*$, called the critical synchronous
and asynchronous stability multipliers, respectively:
\begin{equation}
\lambda_0^* = \lim_{n \rightarrow \infty} {\lambda_{0,n}},~~~
\lambda_1^* = \lim_{n \rightarrow \infty} {\lambda_{1,n}}.
\label{eq:CSM}
\end{equation}
Here $\lambda_0^*$ is just the critical stability multiplier
$\lambda^*$ for the 1D case, and the coupling affects only
$\lambda_1^*$. The values of $\lambda_1^*$ depend on the fixed points,
as shown below.

The invariance of a fixed map $T^*$ under iterations of the period
$p$-tupling renormalization transformation $\cal N$ implies that, if
$T^*$ has a periodic point $(x,y)$ with period $p^n$, then
$B^{-1}(x,y)$ is a periodic point of
$T^*$ with period $p^{n+1}$. Since rescaling leaves the stability
multipliers unaffected, all synchronous orbits of period $p^n$
$(n=0,1,2,\dots)$ have the same stability multipliers, which are just
the critical stability multipliers. Consequently, they have the values
of the stability multipliers of the fixed point of the fixed map
$T^*$:
\begin{equation}
\lambda_0^* =  f^{*'}({\hat x}), \;\;\;
\lambda_1^* = f^{*'}({\hat x})-2G^*({\hat x}),
\label{eq:1CSM}
\end{equation}
where $\hat x$ is the fixed point of the 1D fixed function [i.e.,
${\hat x}= f^* ({\hat x})]$, and $\lambda^*_0$ is just the
critical stability multiplier $\lambda^*$ of the uncoupled 1D map.
Note that $\lambda_1^*$ depends on the reduced coupling fixed function
$G^*(x)$:
\begin{mathletters}
\label{eq:2CSM}
\begin{eqnarray}
\lambda_1^* &=& \lambda^* ~~~{\rm for}~G^*(x) = 0, \label{eq:sma} \\
\lambda_1^* &=& 0 ~~~{\rm for}~G^*(x) = {1 \over 2} f^{*'}(x),
\label{eq:smb} \\
\lambda_1^* &=& 1~~~{\rm for}~G^*(x) = {1 \over 2} [f^{*'}(x) - 1],
\label{eq:smc} \\
\lambda_1^*&=&-1~~~{\rm for}~G^*(x) = {1 \over 2} [f^{*'}(x) + 1],
\label{eq:smd} \\
\lambda_1^*&=&-\lambda^*~~~{\rm for}~G^*(x) = f^{*'}(x),
\label{eq:sme}
\end{eqnarray}
\end{mathletters}
\noindent
where the cases (\ref{eq:sma}) - (\ref{eq:smc}) exist for all $p$, but
the cases (\ref{eq:smd}) - (\ref{eq:sme}) exist only for odd $p$.

Consider a pair of functions $(f^*,{\tilde G})$. Here ${\tilde G}(x)$
is not necessarily a reduced coupling fixed function $G^*(x)$.
When $f_n(x)=f^*(x)$ and $G_n(x)={\tilde G}(x)$, the function
$F_{n,2}^{(p)}({x \over \alpha})$ of Eq.~(\ref{eq:RRE}) will be
denoted by ${\tilde F}_2^{(p)}({x \over \alpha})$. We now examine the
evolution of a pair of functions $(f^* + h, {\tilde G} + \Phi)$ close
to $(f^*,{\tilde G})$ under the reduced renormalization operator
$\tilde {\cal R}$. Linearizing $\tilde {\cal R}$ at
$(f^*,{\tilde G})$, we obtain a linearized
operator $\tilde{\cal L}$ of transforming a pair of infinitesimal
perturbations $(h,\Phi)$:
\begin{equation}
\left( 
\begin{array}{c}
{h_{n+1}} \\ 
{\Phi _{n+1}}
\end{array}
\right) = {\tilde{\cal L}}  \left(
\begin{array}{c}
{h_n} \\ 
{\Phi_n}
\end{array}
\right) = \left( 
\begin{array}{cc}
{\tilde{\cal L}}_1 & \;\;\; 0 \\
{\tilde{\cal L}}_3 & \;\;\; {\tilde{\cal L}}_2
\end{array}
\right) \; \left( 
\begin{array}{c}
{h_n} \\ 
{\Phi_n}
\end{array}
\right),  \label{eq:LOL}
\end{equation}
where
\widetext
\begin{eqnarray}
&h_{n+1}(x) &= [{\tilde{\cal L}}_1 h_n](x) \nonumber \\
&&= {\alpha} \, \delta f_n^{(p)} ({\frac {x} {\alpha}})
\equiv {\alpha}\,
[f_n^{(p)}({\frac {x} {\alpha}}) - f^{*(p)}({\frac {x}
{\alpha}})]_{{\rm linear}} \nonumber \\
&&= {\alpha} {f^{*}}^{\prime}(f^{*(p-1)}({\frac {x} {\alpha}})) \,
\delta f_n^{(p-1)} ({\frac {x} {\alpha}}) + {\alpha} h_n(f^{*(p-1)}
({\frac {x} {\alpha}})), \\
&{\Phi}_{n+1}(x) &= [{\tilde{\cal L}}_2 {\Phi}_n](x) +
[{\tilde{\cal L}}_3 h_n](x) \nonumber \\
&&= {\delta} F_{n,2}^{(p)} ({\frac {x} {\alpha}})
\equiv [F_{n,2}^{(p)}({x \over \alpha}) - {\tilde F}_2^{(p)}
({x \over \alpha})]_{\rm linear},  \\
&[{\tilde{\cal L}}_2 {\Phi}_n](x) & =
[f^{*'}(f^{*(p-1)}({x \over \alpha})) - 2 {\tilde G}(f^{*(p-1)}({x
\over \alpha}))] \, \delta F_{n,2}^{(p-1)}({x \over \alpha})
\nonumber \\
&&~~+ [f^{*(p-1)'}({x \over \alpha}) - 2 {\tilde F}_2^{(p-1)}
({x \over \alpha})] {\Phi_n}(f^{*(p-1)}({x \over \alpha})),
\label{eq:L2} \\
&[{\tilde{\cal L}}_3 {h}_n](x) & =
[{\tilde F}_2^{(p-1)}({x \over \alpha}) f^{*''}(f^{*(p-1)}
({x \over \alpha})) - 2 {\tilde F}_2^{(p-1)}({x \over \alpha})
{\tilde G}'(f^{*(p-1)}({x \over \alpha})) \nonumber \\
&&~~+ f^{*(p-1)'}({x \over \alpha}) {\tilde G}'(f^{*(p-1)}
({x \over \alpha}))] \delta f_n^{(p-1)'}({x \over \alpha}) +
{\tilde F}_2^{(p-1)}({x \over \alpha}) h'_n(f^{*(p-1)}({x \over
\alpha})) \nonumber \\
&&~~+ {\tilde G}(f^{*(p-1)}({x \over \alpha})) \delta
f_n^{(p-1)'}({x \over \alpha}).
\end{eqnarray}
\narrowtext
\noindent
Here the variations $\delta f_n^{(p)} ({\frac {x} {\alpha}})$ and
${\delta} F_{n,2}^{(p)}({\frac {x} {\alpha}})$ are introduced as the
linear terms (denoted by $[f_n^{(p)}({\frac {x} {\alpha}}) - f^{*(p)}
({\frac {x} {\alpha}})]_{{\rm linear}}$ and $[F_{n,2}^{(p)} ({\frac
{x} {\alpha}}) - {\tilde{F}}_2^{(p)} ({\frac {x} {\alpha}})]_{{\rm
linear}}$) in $h$ and $\Phi$ of the deviations of $f_n^{(p)} ({\frac
{x} {\alpha}})$ and $F_{n,2}^{(p)} ({\frac {x} {\alpha}})$
from $f^{*(p)}({x \over \alpha})$ and ${\tilde{F}}_2^{(p)}({x \over
\alpha})$, respectively.

When ${\tilde G}(x)$ is a reduced fixed coupling function $G^*(x)$ of
Eq.~(\ref{eq:RCFFTN}), the operator $\tilde{\cal L}$ of
Eq.~(\ref{eq:LOL}) becomes a linearized transformation of
$\tilde{\cal R}$ at a fixed point $(f^*,G^*)$. A pair of perturbations
$(h^*,\Phi^*)$ is then called an eigenperturbation with eigenvalue
$\nu$, if it satisfies
\begin{equation}
\nu \, \left( 
\begin{array}{c}
{h^*} \\
{\Phi^*}
\end{array}
\right) = {\cal L} \left( 
\begin{array}{c}
{h^*} \\
{\Phi^*}
\end{array}
\right).
\end{equation}
The reducibility of $\tilde{\cal L}$ into a semiblock form implies
that to determine the eigenvalues of $\tilde{\cal L}$ it is sufficient
to work independently in each of $h(x)$ subspace and $\Phi(x)$
subspace. That is, one can find eigenvalues of ${\tilde{\cal L}}_1$
and ${\tilde{\cal L}}_2$ separately and then they give the whole
spectrum of $\cal L$.

We first solve the eigenvalue equation for ${\tilde{\cal L}}_1$, i.e.,
\begin{equation}
\nu h^*(x) = [{\tilde{\cal L}}_1 h^*](x).
\end{equation}
Note that this is just the eigenvalue equation for the 1D map case.
It has been shown that there exists only one relevant eigenvalue
$\delta$, associated with scaling of the nonlinearity parameter,
whose values vary depending on $p$
\cite{Derrida,Hu,Eckmann,Hao,Hauser,Chang,Kenny,Urumov}. However, note
that although the eigenvalue $\delta$ of ${\tilde{\cal L}}_1$
is also an eigenvalue of $\tilde{\cal L}$, $(h,0)$ is not an
eigenperturbation of $\tilde{\cal L}$ unless ${\tilde{\cal L}}_3$ is a
null operator.

We next consider a perturbation of the form $(0,\Phi)$ having only the
coupling part. In this case $(0,\Phi^*)$ can be an eigenperturbation
of $\tilde{\cal L}$, only if $\Phi^*(x)$ satisfies
\begin{equation}
\nu {\Phi^*}(x) = [{\tilde{\cal L}}_2 {\Phi^*}](x).
\label{eq:CEVEQ}
\end{equation}
Eigenvalues associated with coupling perturbations are called CE's.

For the case of coupling perturbation $(0,\Phi)$ to
$(f^*,{\tilde G})$, ${\tilde {\cal L}}_2$ of Eq.~(\ref{eq:L2})
becomes:
\begin{eqnarray}
[{\tilde{\cal L}}_2 {\Phi}](x)
&=& {\delta} F_{2}^{(p)} ({\frac {x} {\alpha}}) \nonumber \\
&=&
[f^{*'}(f^{*(p-1)}({x \over \alpha})) - 2 {\tilde G}(f^{*(p-1)}({x
\over \alpha}))]  \delta F_{2}^{(p-1)}({x \over \alpha})
\nonumber \\
&& + [f^{*(p-1)'}({x \over \alpha}) - 2 {\tilde F}_2^{(p-1)}
({x \over \alpha})]
{\Phi}(f^{*(p-1)}({x \over \alpha})).
\label{eq:dF2p}
\end{eqnarray}
We first consider the lowest two cases with $p=2,3$. For the
period-doubling case, $\delta F_2^{(2)}({x \over \alpha})$ becomes
\widetext
\begin{mathletters}
\label{eq:p2df2}
\begin{eqnarray}
\delta F_{2}^{(2)} ({x \over \alpha}) &=&
f^{*'}(f^*({x \over \alpha})) {\Phi}({x \over \alpha})
+ f^{*'}({x \over \alpha}) {\Phi}(f^*({x \over \alpha}))
~~{\rm for}~{\tilde G}(x)=0, \label{eq:p2a} \\
\delta F_{2}^{(2)} ({x \over \alpha}) &=& 0
 ~~~{\rm for}~{\tilde G}(x)= {1 \over 2} f^{*'}(x), \label{eq:p2b} \\
\delta F_{2}^{(2)} ({x \over \alpha}) &=&
{\Phi}({x \over \alpha}) + {\Phi}(f^{*}({x \over \alpha}))
~~~{\rm for}~{\tilde G}(x)= {1 \over 2} [f^{*'}(x)-1],
\label{eq:p2c}\\
\delta F_{2}^{(2)} ({x \over \alpha}) &=&
-{\Phi}({x \over \alpha}) - {\Phi}(f^{*}({x \over \alpha}))
~~~{\rm for}~{\tilde G}(x)= {1 \over 2} [f^{*'}(x)+1],
\label{eq:p2d} \\
\delta F_{2}^{(2)} ({x \over \alpha}) &=&
-f^{*'}(f^*({x \over \alpha})) {\Phi}({x \over \alpha})
-f^{*'}({x \over \alpha}) {\Phi}(f^*({x \over \alpha}))
       ~~~{\rm for}~{\tilde G}(x)= f^{*'}(x). \label{eq:p2e}
\end{eqnarray}
\end{mathletters}
\noindent
Note that only ${\tilde G}$'s for the first three cases
(\ref{eq:p2a}) - (\ref{eq:p2c}) are the reduced coupling fixed
functions $G^*$'s.
Using Eqs.~(\ref{eq:p2F2})
and (\ref{eq:p2df2}), we also find
$\delta F_2^{(3)}({x \over \alpha})$ for $p=3$,
\begin{mathletters}
\label{eq:p3df2}
\begin{eqnarray}
\delta F_{2}^{(3)} ({x \over \alpha}) &=&
\sum_{i=0}^{2} f^{*(i)'}({x \over \alpha}) \Phi(f^{*(i)}({x
\over \alpha})) f^{*(2-i)'}(f^{*(i+1)}({x \over \alpha}))
~~{\rm for}~{\tilde G}(x)=0, \label{eq:p3a} \\
\delta F_{2}^{(3)} ({x \over \alpha}) &=& 0
   ~~~{\rm for}~{\tilde G}(x)= {1 \over 2} f^{*'}(x),
   \label{eq:p3b} \\
\delta F_{2}^{(3)} ({x \over \alpha}) &=&
{\sum_{i=0}^2} {\Phi}(f^{*(i)}({x \over \alpha}))
  ~~~{\rm for}~{\tilde G}(x)= {1 \over 2} [f^{*'}(x)-1],
  \label{eq:p3c}\\
\delta F_{2}^{(3)} ({x \over \alpha}) &=&
          {\sum_{i=0}^2} {\Phi}(f^{*(i)}({x \over \alpha}))
 ~~~{\rm for}~{\tilde G}(x)= {1 \over 2} [f^{*'}(x)+1],
 \label{eq:p3d} \\
\delta F_{2}^{(3)} ({x \over \alpha}) &=&
\sum_{i=0}^{2} f^{*(i)'}({x \over \alpha}) \Phi(f^{*(i)}
({x \over \alpha})) f^{*(2-i)'}(f^{*(i+1)}({x \over \alpha}))
       ~~~{\rm for}~{\tilde G}(x)= f^{*'}(x), \label{eq:p3e}
\end{eqnarray}
\end{mathletters}
\noindent
where $f^{*(0)}(x)=x$.
For this period-tripling case, all $\tilde G$'s in
Eqs.~(\ref{eq:p3a}) - (\ref{eq:p3e}) are the reduced coupling fixed
functions $G^*$'s.

We obtain $\delta F_2^{(p)} ({x \over \alpha})$ for the general period
$p$-tupling case by induction. Suppose that $\delta F_2^{(p)}
({x \over \alpha})$ for the period $p$-tupling case is
\begin{mathletters}
\label{eq:pndF2}
\begin{eqnarray}
\delta F_{2}^{(p)} ({x \over \alpha}) &=&
\sum_{i=0}^{p-1} f^{*(i)'}({x \over \alpha}) \Phi(f^{*(i)}
({x \over \alpha}))
f^{*(p-1-i)'}(f^{*(i+1)}({x \over \alpha}))
~~~{\rm for}~{\tilde G}(x)=0, \\
\delta F_{2}^{(p)} ({x \over \alpha}) &=& 0
  ~~~{\rm for}~{\tilde G}(x)= {1 \over 2} f^{*'}(x), \\
\delta F_{2}^{(p)} ({x \over \alpha}) &=&
{\sum_{i=0}^{p-1}} {\Phi}(f^{*(i)}({x \over \alpha}))
~~~{\rm for}~{\tilde G}(x)= {1 \over 2} [f^{*'}(x)-1], \\
\delta F_{2}^{(p)} ({x \over \alpha}) &=&
 \left \{
 \begin{array}{cl}
- \displaystyle{{\sum_{i=0}^{p-1}} {\Phi}(f^{*(i)}({x \over \alpha}))}
~&{\rm for~even~} p  \\
\displaystyle{  {\sum_{i=0}^{p-1}} {\Phi}(f^{*(i)}({x \over \alpha}))}
  ~&{\rm for~odd~} p
 \end{array}
 \right.
 ~{\rm for}~{\tilde G}(x)= {1 \over 2} [f^{*'}(x)+1], \\
\delta F_{2}^{(p)} ({x \over \alpha}) &=&
  \left \{
  \begin{array}{cl}
- \displaystyle{ \sum_{i=0}^{p-1} f^{*(i)'}({x \over \alpha})
\Phi(f^{*(i)}({x \over \alpha})) f^{*(p-1-i)'}(f^{*(i+1)}
({x \over \alpha}))}~&{\rm for~even~} p \\
 \displaystyle{\sum_{i=0}^{p-1} f^{*(i)'}({x \over \alpha})
\Phi(f^{*(i)}({x \over \alpha})) f^{*(p-1-i)'}(f^{*(i+1)}({x \over
\alpha}))}~&{\rm for~odd~} p
  \end{array}
\right. \nonumber \\
&& {\rm for}~ {\tilde G}(x)=f^{*'}(x).
\end{eqnarray}
\end{mathletters}
\narrowtext
\noindent
For the case of the next period $(p+1)$-tupling, $\delta F_2^{(p+1)}$
of Eq.~(\ref{eq:dF2p}) becomes
\begin{eqnarray}
\delta F_2^{(p+1)} &=&
[f^{*'}(f^{*(p)}({x \over \alpha})) - 2 {\tilde G}(f^{*(p)}({x \over
\alpha}))] \, \delta F_{2}^{(p)}({x \over \alpha}) \nonumber \\
&&~+ [f^{*(p)'}({x \over \alpha}) - 2 {\tilde F}_2^{(p)}
({x \over \alpha})] {\Phi}(f^{*(p)}({x \over \alpha})).
\end{eqnarray}
It follows from Eqs.~(\ref{eq:pnF2}) and (\ref{eq:pndF2}) that
$\delta F_2^{(p+1)}$ for the period $(p+1)$-tupling case also
satisfies Eq.~(\ref{eq:pndF2}). Note that the lowest even and odd
cases, $p=2$ and $3$, satisfy the assumption (\ref{eq:pndF2})
[see Eqs.~(\ref{eq:p2df2}) and (\ref{eq:p3df2})].
The equation (\ref{eq:pndF2}) is therefore valid for all $p$.

We now find CE's associated with coupling perturbations.
For the zero-coupling case of $G^*(x)=0$, the CE equation
(\ref{eq:CEVEQ}) becomes
\begin{eqnarray}
\nu {\Phi^*}(x) &=& [{\tilde{\cal L}}_2 {\Phi}^*] (x)
= \delta F_2^{(p)}({x \over \alpha}) \nonumber \\
 &=&
\sum_{i=0}^{p-1} f^{*(i)'}({x \over \alpha}) \Phi^*(f^{*(i)}({x \over
\alpha})) f^{*(p-1-i)'}(f^{*(i+1)}({x \over \alpha})).
\label{eq:zceveq}
\end{eqnarray}
Using the fact that ${f^*}^{\prime}(0)=0$, it can be easily shown that
when $x=0$, the equation (\ref{eq:zceveq}) becomes
\begin{equation}
\nu\, {\Phi^*}(0) = [ {\prod_{i=1}^{p-1}} {f^*}^{\prime}
(f^{*(i)}(0))]\, {\Phi^*}(0).  \label{eq:Phi0}
\end{equation}
Letting $x \rightarrow 0$ in Eq.~(\ref{eq:1DD}), we also have
\begin{equation}
{\prod_{i=1}^{p-1}} {f^*}^{\prime}(f^{*(i)}(0)) = {\lim_{x
\rightarrow 0}} { \frac{{f^*}^{\prime}(x) }{{f^*}^{\prime}(
{\frac{x }{\alpha}})}} =\alpha.
\end{equation}
Then Eq.~(\ref{eq:Phi0}) reduces to 
\begin{equation}
\nu \, {\Phi^*}(0) = \alpha {\Phi^*}(0).
\end{equation}
There are two cases.
For the case $\Phi^*(0) \neq 0$, we have the first CE
\begin{equation}
\nu_1=\alpha.  \label{eq:1stCE}
\end{equation}
The eigenfunction $\Phi^*_1(x)$ with CE $\nu_1$ is of the form,
\begin{equation}
\Phi^*_1(x) = 1 + a_1 x + a_2 x^2 +\cdots,
\end{equation}
where $a_i$'s $(i=1,2,\dots)$ are some constants. For the other case
$\Phi^*(0) = 0$, it is found that ${f^*}^{\prime}(x)$ is an
eigenfunction for the CE equation (\ref{eq:zceveq}).
When $\Phi^*(x) = f^{*'}(x)$, the equation (\ref{eq:zceveq}) becomes
\begin{equation}
\nu {f^*}^{\prime}(x) = p f^{*(p)'}({x \over \alpha}) = p
{f^*}^{\prime}(x).
\end{equation}
We therefore have the second relevant CE 
\begin{equation}
\nu_2 = p,
\end{equation}
with reduced coupling eigenfunction
\begin{equation}
\Phi^*_2(x) = {f^*}^{\prime}(x).
\end{equation}
Note that $\Phi_2^*(x)$ has no constant term, while $\Phi_1^*(x)$ has
a constant term. Thus we find two relevant CE's, $\nu_1 = \alpha$ and
$\nu_2 = p$, for the zero-coupling case.

The $n$th image $\Phi_n$ of a general reduced coupling perturbation
$\Phi$ under the linear transformation ${\tilde{\cal L}}_2$ has the
form,
\begin{eqnarray}
\Phi_n(x) &=& [{\tilde{\cal L}}_2^n \Phi](x) \nonumber \\
& \sim & {\alpha_1} {\nu_1^n} {\Phi_1^*}(x) +
{\alpha_2} {\nu_2^n} {\Phi_2^*}(x)~~~{\rm for~large}~n,
\label{eq:Phin}
\end{eqnarray}
because the irrelevant part of $\Phi_n$ becomes negligibly small for
large $n$. Here $\alpha_1$ and $\alpha_2$ are relevant components.

A coupling is called linear or nonlinear according to its leading
term. In the case of a linear coupling, in which the coupling
perturbation $\varphi(x,y)$ has a leading linear term, its reduced
coupling function $\Phi(x)$ has a leading constant term, and hence
$\Phi(0) \neq 0$.
However, for any other case of nonlinear coupling with a leading
nonlinear term, its reduced coupling function has no constant term,
and hence $\Phi(0)=0$. Note that the relevant component $\alpha_1$
becomes
zero for the nonlinear-coupling case, while it is nonzero for the
linear-coupling case. Consequently, the CB associated with
coupling perturbations is governed by two relevant CE's $\nu_1=\alpha$
and $\nu_2=p$ for the linear-coupling case, but by only one CE
$\nu_2 = p$ for the nonlinear-coupling case, which will be confirmed
in Sec.~{\ref{sec:EX}.

We now consider the cases of two (four) other reduced coupling fixed
functions $G^*(x)$'s for even (odd) $p$ [see Eq.~(\ref{eq:RCFFTN})].
They are associated with CB at critical points other than the
zero-coupling critical point, as will be seen in Sec.~{\ref{sec:EX}.
Since $\delta F_2^{(p)}({x \over \alpha}) =0$ for
$G^*(x)= {1 \over 2} f^{*'}(x)$, ${\tilde {\cal L}}_2$ becomes a null
operator. Hence there exist no CE's, and the CB is essentially
the same as that for the 1D case.
When $G^*(x) = {1 \over 2} [f^{*'}(x) -1]$, the CE
equation (\ref{eq:CEVEQ}) becomes
\begin{eqnarray}
\nu {\Phi^*}(x) &=& [{\tilde{\cal L}}_2 {\Phi^*}] (x)
= \delta F_2^{(p)}({x \over \alpha}) \nonumber \\
 &=&
\sum_{i=0}^{p-1} {\Phi^*}(f^{*(i)}({x \over \alpha})).
\label{eq:3ceveq}
\end{eqnarray}
There exists a relevant CE
\begin{equation}
\nu=p,
\end{equation}
when $\Phi^*(x)$ is a nonzero constant
function, i.e., $\Phi^*(x)=b$ ($b$ is a nonzero constant).
For odd $p$ there are two additional reduced coupling fixed functions,
$G^*(x) = {1 \over 2} [f^{*'}(x)+1]$ and $G^*(x) = f^{*'}(x)$.
The CE equation (\ref{eq:CEVEQ}) for $G^*(x) = {1 \over 2}
[f^{*'}(x)+1]$ is just that of Eq.~(\ref{eq:3ceveq}). Therefore it has
the same CE $\nu=p$
as that for the case $G^*(x) = {1 \over 2} [f^{*'}(x)-1]$.
However, the critical asynchronous stability multipliers $\lambda_1^*$
for the two cases are different [see eq.~({\ref{eq:2CSM})]. When
$G^*(x) = f^{*'}(x)$, the CE equation (\ref{eq:CEVEQ}) is the same as
that for the case $G^*(x)=0$. Hence it also has two relevant CE's
$\nu_1 = \alpha$ and $\nu_2 = p$. However, $\lambda_1^*$ for this case
is different from that for the case $G^*(x)=0$, as can be seen in
Eq.~(\ref{eq:2CSM}). The results of relevant CE's, along with those of
the critical asynchronous stability multipliers, are listed in Table
\ref{table1}.

\section{Linear and nonlinear couplings}
\label{sec:EX}

We choose $f(x) = 1 - A x^2$ as the uncoupled 1D map and consider two
kinds of couplings, linear and nonlinear couplings. As examples of the
linear- and nonlinear-coupling cases, we study the
linearly- and dissipatively-coupled maps, respectively, and confirm
the renormalization results. The structure of the critical set (set
of the critical points) varies depending on the kind of coupling.
In the linearly-coupled case, an infinite number of critical line
segments, together with the zero-coupling critical point, constitute
the critical set, whereas in the dissipative case, the critical set
consists of the only one critical line segment, one end of which is
the zero-coupling critical point. The CB also depends on the position
on the critical set. For even (odd) $p$, three (four) kinds of fixed
points govern the CB for the linearly-coupled case, while only two
(three) kinds of fixed points govern the CB for the dissipative case.

\subsection{Linearly coupled maps}
\label{sub:LCM}

We numerically study the CB of period $p$-tuplings
in two linearly-coupled 1D maps with the coupling
function
\begin{equation}
g(x,y) = {c \over 2} (y-x),
\label{eq:TLC}
\end{equation}
where $c$ is the coupling parameter.
The critical scaling behavior depends on whether $p$ is even or odd.

As an example of odd period $p$-tuplings, we take the period-tripling
case $(p=3)$ and study its CB. The stability diagrams of
synchronous orbits with period $q=3^n$ $(n=0,1,2,3)$ are shown in
Figs.~\ref{SDT1}-\ref{SDT2}. As noted in Sec.~\ref{sub:TCM}, the
stable region of each synchronous orbit of level $n$ (period $3^n$)
in the parameter
plane is bounded by the four synchronous and asynchronous bifurcation
curves determined by the equations $\lambda_{i,n}= \pm 1$ for $i=0,1$.
Since each synchronous orbit of level $n$ is born by its own
synchronous saddle-node bifurcation (which occurs for
$\lambda_{0,n}=1$), a sequence of stability regions with increasing
$n$ is not connected, unlike the period-doubling case \cite{Kim1}.
We now examine the treelike structure of stability regions.
Figure \ref{SDT1}
shows the stability regions of synchronous orbits of the lowest two
levels (i.e., $n=0$ and $1$). The synchronous orbit with period $q=1$
is stable in some quadrilateral-shape region containing the $c=0$ line
segment. However, the stability region of the next level $n=1$
consists of three quadrilateral-shape areas.  They can be regarded as
``daughter'' quadrilaterals of the ``mother'' quadrilateral of level
$0$. That is, it may be thought that they branch off from the mother
quadrilateral.

We next consider the stability regions of higher levels
in Fig.~\ref{SDT2}. The branchings occur from the central one
containing the $c=0$ line segment and its nearest neighboring one
(i.e., the right one) among the three quadrilaterals of level $1$
[see Figs.~\ref{SDT2}(a) and \ref{SDT2}(b)], while there is no
branching from the left one [see Fig.~\ref{SDT2}(c)]. This rule
governs the treelike structure of the stability regions. That is, for
each level $n$ branchings occur only from two quadrilateral-shape
areas, the central one containing the $c=0$ line segment and its
nearest neighboring one. However, an infinite number of successive
branchings occur only for the case of the central quadrilateral
including the $c=0$ line segment [see Fig.~\ref{SDT2}(a)]. For the
case of the nearest neighboring quadrilateral, branching occurs only
once, and after that, successive quadrilaterals of higher levels pile
up without any further branchings [e.g., see Fig.~\ref{SDT2}(b)].
For the cases of quadrilaterals other than the central and its nearest
neighboring ones, successive quadrilaterals of higher levels pile up
without any branchings [e.g., see Fig.~\ref{SDT2}(c)].

A sequence of stability regions with increasing period is called a
``period-tripling route'', like the period-doubling case \cite{Kim1}.
There are two kinds of period-tripling route.
The sequence of the quadrilateral-shape areas containing the $c=0$
line segment converges to the zero-coupling point $c=0$ on the
$A=A_{\infty}^{(3)}$ line, where $A_\infty^{(3)}$
$(=1.786\, 440\, 255\, 563\, 639\, 354\, 534\, 447 \dots)$ is
the accumulation point of the period-tripling sequence for the 1D
case. It will be referred to as the $Z_3$ route.
On the other hand, a sequence of quadrilaterals which piles up without
successive branchings converges to a critical line. For example, the
leftmost one is the line joining two points $c_l$
$(=-3.590\, 291\, 636\, 032\, 974\, 400\, 442 \dots)$ and $c_r$
$(=-3.482\, 633\, 674\, 606\, 564\, 177\, 473 \dots)$ on the
$A=A_\infty^{(3)}$ line [see Fig.~\ref{SDT2}(c)].
This kind of route will be called an $L_3$ route. Note that there are
infinitely many $L_3$ routes, while the $Z_3$ route converging to the
zero-coupling critical point $(A_\infty, 0)$ is unique. Hence an
infinite number of critical line segments, together with the
zero-coupling critical point, constitute the critical set.

We now study the critical scaling behavior on the critical set. First,
consider the case of the $Z_3$ route ending at the zero-coupling
critical point. The CB for this zero-coupling case is governed by the
zero-coupling fixed map (\ref{eq:FM}) with $g^*(x,y)=0$, which has two
relevant CE's $\nu_1=\alpha$ $(=-9.277\,341 \dots)$ and $\nu_2=3$
(see Table \ref{table1}).

The quadrilateral-shape areas of the levels $n=1,2,3$ in the $Z_3$
route are shown in Fig.~\ref{PS}. Note that
Figs.~\ref{PS}(a), \ref{PS}(b), and \ref{PS}(c) nearly coincide except
for small numerical differences. It is therefore expected that the
height and width $h_n$ and $w_n$ \cite{note2} of the
quadrilateral-shape area of level $n$ may geometrically contract,
\begin{equation}
h_n \sim \delta^{-n},\;\;w_n \sim \alpha^{-n}\;\;{\rm for\;large}\;n,
\end{equation}
where $\delta = 55.247\,026\dots$~.
Thus the scaling factors of the nonlinearity and coupling parameters
become $\delta$ and $\alpha$, respectively, which will be shown
explicitly below.

We follow the synchronous orbits of period $q=3^n$ up to level $n=10$
in the $Z_3$ route, and obtain a self-similar sequence of parameters
$(A_n,c_n)$, at which each orbit of level $n$ has some given stability
multipliers $(\lambda_0,\lambda_1)$ (e.g., $\lambda_0=-1$ and
$\lambda_1=1$). Then the sequence $\{ (A_n,c_n) \}$ converges
geometrically to the zero-coupling critical point
$(A_\infty^{(3)},0)$. In order to see the convergence of each of the
two scalar sequences $\{ A_n \}$ and $\{ c_n \}$,
define $\delta_n \equiv { {\Delta A_n} \over {\Delta A_{n+1}} }$ and
$\mu_n \equiv { {\Delta c_n} \over {\Delta c_{n+1}} }$,
where $\Delta A_n = A_n - A_{n-1}$ and $\Delta c_n = c_n - c_{n-1}$.
Then they converge to their limit values $\delta$ and $\mu$,
respectively, as shown in Table \ref{table2}. Hence the two sequences
$\{ A_n \}$ and $\{ c_n \}$ obey one-term scaling laws asymptotically:
\begin{equation}
\Delta A_n \sim \delta^{-n},\;\;\;
\Delta c_n  \sim \mu^{-n}\;\;\;{\rm for\;large\;}n,
\label{eq:OTSL}
\end{equation}
where  $\delta=55.247 \dots$ and $\mu=-9.277 \dots$~. Note that the
nonlinearity-parameter scaling factor $\delta$ is just that for the
1D case, and the value of the coupling-parameter scaling factor $\mu$
is close to that of the first CE $\nu_1$ $(=\alpha)$.

In order to take into account the effect of the second relevant CE
$\nu_2$ $(=3)$ on the scaling of the sequence $\{ \Delta c_n \}$,
we extend the simple one-term scaling law (\ref{eq:OTSL}) to a
two-term scaling law \cite{Kim3}:
\begin{equation}
\Delta c_n \sim C_1 \mu_{1}^{-n} + C_2 \mu_{2}^{-n}
\;\;\;{\rm for\;large\;}n,
\label{eq:TTSL1}
\end{equation}
where $| \mu_2 | > | \mu_1 | $, and $C_1$ and $C_2$ are some
constants. This is a kind of multiple scaling law \cite{MR}.
Eq.~(\ref{eq:TTSL1}) gives
\begin{equation}
\Delta c_n = s_1 \Delta c_{n+1} -s_2 \Delta c_{n+2},
\label{eq:RE}
\end{equation}
where $s_1 = \mu_1 + \mu_2$ and $s_2 = \mu_1 \mu_2$.
Then $\mu_1$ and $\mu_2$ are solutions of the following quadratic
equation,
\begin{equation}
\mu^2 - s_1 \mu + s_2 =0.
\label{eq:QE}
\end{equation}
To evaluate $\mu_1$ and $\mu_2$, we first obtain $s_1$ and $s_2$ from
$\Delta c_n$'s using Eq.~(\ref{eq:RE}):
\begin{equation}
s_1 = { {\Delta c_n \Delta c_{n+1} - \Delta c_{n-1} \Delta c_{n+2}}
\over {\Delta c_{n+1}^2 - \Delta c_n \Delta c_{n+2}} }, \;\;
s_2 = { {\Delta c_n^2 - \Delta c_{n+1} \Delta c_{n-1}}
\over {\Delta c_{n+1}^2 - \Delta c_n \Delta c_{n+2}} }.
\label{eq:T1T2}
\end{equation}
Note that Eqs.~(\ref{eq:TTSL1})-(\ref{eq:T1T2}) hold only for large
$n$. In fact the values of $s_i$'s and $\mu_i$'s $(i=1,2)$ depend on
the level $n$. Therefore we explicitly denote $s_i$'s and $\mu_i$'s by
$s_{i,n}$'s and $\mu_{i,n}$'s, respectively. Then each of them
converges to a constant as $n \rightarrow \infty$:
\begin{equation}
\lim_{n \rightarrow \infty} s_{i,n} = s_i, \;\;\;
\lim_{n \rightarrow \infty} \mu_{i,n} = \mu_i,\;\;i=1,2.
\end{equation}

Three sequences $\{ \mu_{1,n} \}$, $\{ \mu_{2,n} \}$, and
$\displaystyle{ \{ {\mu_{1,n}^2 / \mu_{2,n}} \} }$ are shown in
Table \ref{table3}.
The second column shows rapid convergence of $\mu_{1,n}$ to its limit
value $\mu_1$ $(=-9.277 \, 341 \dots)$, which is close to
the renormalization result of the first relevant CE $\nu_1$
($=\alpha$). (Its convergence to $\alpha$ is faster than that for
the case of the above one-term scaling law.)
From the third and fourth columns, we also find that the second
scaling factor $\mu_2$ is given by a product of two relevant CE's
$\nu_1$ and $\nu_2$,
\begin{equation}
\mu_2 = {\nu_1^2 \over \nu_2},
\end{equation}
where $\nu_1=\alpha$ and $\nu_2=3$.
It has been known that every scaling factor in the multiple-scaling
expansion of a parameter is expressed by a product of the eigenvalues
of a linearized renormalization operator \cite{MR}.

We also study the effect of CE's on the asynchronous stability
multipliers of synchronous periodic orbits. Consider the
two-coupled map (\ref{eq:CM})
with $f(x)=f_c(x)$ and $g(x,y)= \varepsilon \varphi(x,y)$.
Here $f_c(x)$ is the 1D critical map with the nonlinearity parameter
set to its critical value $A=A_\infty^{(p)}$ and $\varepsilon$ is an
infinitesimal coupling parameter.
The map for $\varepsilon = 0$ is just the critical map $T_c$ at the
zero-coupling critical point consisting of two uncoupled 1D
critical maps $f_c$. It is attracted to the zero-coupling fixed map
(\ref{eq:FM}) with $F^*(x,y) = f^*(x)$ under iterations of the
renormalization transformation ${\cal N}$ of Eq.\ (\ref{eq:RON}).
Hence the reduced coupling function $G(x)$ $[= \varepsilon \Phi \equiv
\varepsilon\partial \varphi(x,y) / \partial y |_{y=x}]$ corresponds to
an infinitesimal reduced coupling perturbation to the reduced coupling
fixed function $G^*(x)=0$.

In the period $p$-tupling case, the stability multipliers
$\lambda_{0,n}$ and $\lambda_{1,n}$ of the $p^n$-periodic orbit are
the same as those of the fixed point of the $n$ times renormalized map
${\cal N}^n (T)$, which are given by
\begin{equation}
\lambda_{0,n} = {f'_n}({\hat x}_n), \;\;
\lambda_{1,n} = {f'_n}({\hat x}_n)- 2 {G_n}({\hat x}_n)
              \simeq  {f'_n}({\hat x}_n)- 2 \varepsilon
              \Phi({\hat x}_n).
\label{eq:MULTI2}
\end{equation}
Here $(f_n,G_n)$ be the $n$th image of $(f_c,G)$ under the
reduced renormalization transformation $\tilde{\cal R}$,
${\hat x}_n$ is just the fixed point of $f_n(x)$ [i.e.,
$ {\hat x}_n={f_n}({\hat x}_n)$] and converges to the fixed point
$\hat x$ of the 1D fixed map $f^*(x)$ as $n \rightarrow \infty$. The
first stability multiplier $\lambda_{0,n}$ converges to the 1D
critical stability multiplier $\lambda^* =f^{*'}({\hat x})$ as
$n \rightarrow \infty$.
For the period-tripling case, $\lambda^* = -1.872\,705\,929 \dots$~.
Since $G_n(x)  \simeq [{\tilde{\cal L}}_2^n G](x) = \varepsilon
\Phi_n(x)$ [$\Phi_n(x)$ is given in Eq.~(\ref{eq:Phin})],
the asynchronous stability multiplier has the form
\begin{eqnarray}
{\lambda_{1,n}} & \simeq &  {\lambda_{0,n}} -2 \varepsilon \Phi_n(x)
\nonumber \\
 &\simeq&  {\lambda^*} + {\varepsilon}
\left[ e_1 {\nu_1^n} + e_2 {\nu_2^n} \right]
\;\;{\rm for\;\;large}\;\;n,
\end{eqnarray}
where ${e_1}=-2{\alpha_1} {\Phi^*_1({\hat x})}$
and ${e_2}=-2{\alpha_2} {f^*}'({\hat x})$.
Therefore the slope $S_n$ of $\lambda_{1,n}$ at the zero-coupling
critical point ($\varepsilon=0$) is
\begin{equation}
S_n \equiv
\left. {\displaystyle {\frac {\partial \lambda_{1,n}}
{\partial \varepsilon} } }\right|_{\varepsilon=0}
\label{eq:slope}
\simeq e_1  {\nu_1^n} + e_2 {\nu_2^n}\;\;
{\rm for\;\;large} \;\;n.
\end{equation}
Here the coefficients $e_1$ and $e_2$ depend on the
initial reduced function $\Phi(x)$, because $\alpha_n$'s are
determined only by $\Phi(x)$. Note that the coefficient $e_1$ is zero
for the nonlinear-coupling case, whereas it is nonzero for the
linear-coupling case. Hence the growth of $S_n$ for large $n$ is
governed by the two relevant CE's $\nu_1 = \alpha$ and $\nu_2 = p$ for
the linear-coupling case, but only by the second
relevant CE $\nu_2 = p$ for the nonlinear-coupling case.

Figure \ref{ASM1} shows three plots of
$\lambda_{1,n}(A_\infty^{(3)},c)$ versus $c$ for $n=2,3,$ and $4$.
The slope $S_n$ of $\lambda_{1,n}$ at the zero-coupling critical point
increases with $n$, and obeys well the two-term scaling law,
\begin{equation}
S_n \sim d_1 r_1^n + d_2 r_2^n \;\;{\rm for \; large\;}n,
\end{equation}
where $d_1$ and $d_2$ are some constants and $|r_1| > |r_2|$. This
equation gives
\begin{equation}
S_{n+2} = t_1 S_{n+1} - t_2 S_{n},
\end{equation}
where $t_1 = r_1 + r_2$ and $t_2 = r_1 r_2$.
As in the scaling for the coupling parameter, we first obtain $t_1$
and $t_2$ of level $n$ from $S_n$'s:
\begin{equation}
t_{1,n} = {{S_{n+1} S_n - S_{n+2} S_{n-1}} \over {S_n^2 - S_{n+1}
S_{n-1}}},\;\;t_{2,n} = {{S_{n+1}^2 - S_n S_{n+2}} \over {S_n^2 -
S_{n+1} S_{n-1}}}.
\end{equation}
Then the scaling factors $r_{1,n}$ and $r_{2,n}$ of level $n$ are
given by the roots of the quadratic equation, $r_n^2 - t_{1,n} r_n
+ t_{2,n} = 0$. They are listed in Table \ref{table4} and converge
to constants $r_1$ $(=\nu_1)$ and $r_2$ $(=\nu_2)$ as
$n \rightarrow \infty$, whose
accuracies are higher than those of the coupling-parameter scaling
factors.

We next consider the cases of $L_3$ routes, each of which converges to
a critical line segment. In each $L_3$ route, there are two kinds of
self-similar sequences of parameters $(A_n,c_n)$, at which each orbit
of level $n$ has some given stability multipliers; the one convergs to
the left end point of the critical line segment and the other convergs
to the right end point. As an example, consider the leftmost $L_3$
route [see Fig.~\ref{SDT2}(c)], which converges to the critical line
segment with two ends $(A_\infty^{(3)},c_l)$ and
$(A_\infty^{(3)},c_r)$. We follow, in the leftmost $L_3$ route, two
self-similar sequences of parameters, one converging to the left end
and the other converging to the right end. In both cases, the sequence
$\{ A_n \}$ converges geometrically to its accumulation value
$A_\infty^{(3)}$ as in the case of the $Z_3$ route,
\begin{equation}
\Delta A_n \sim \delta^{-n} \;\;{\rm for\;large}\; n,
\end{equation}
where $\Delta A_n = A_n - A_{n-1}$ and $\delta=55.247\dots$~.
The sequences $\{ c_n \}$ for both cases also obey the one-term
scaling law,
\begin{equation}
\Delta c_n \sim \mu^{-n} \;\;{\rm for\;large}\; n,
\end{equation}
where $\Delta c_n = c_n - c_{n-1}$. The convergence of the scaling
factor $\mu_n$ of level $n$ to its limit value $\mu$ $(=3)$ is
shown in Table \ref{table5}. Note that the value of $\mu$ is
different from that $(\mu=\alpha)$ at the zero-coupling critical
point. Although the scaling factors of the coupling parameter at
both ends are the same, the critical asynchronous stability
multipliers $\lambda_1^*$'s of Eq.~(\ref{eq:CSM}) at both ends
have different values. The convergence of the sequence
$\{ \lambda_{1,n} \}$ to its limit value $\lambda_1^*$ is also
shown in Table \ref{table5}. At the left (right) end, $\lambda_1^*
= 1\;(-1)$. Comparing the values of $\mu$ and $\lambda_1^*$ with
those of the CE $\nu$ and $\lambda_1^*$ listed in
Table \ref{table1}, we find that the CB near the left end is
governed by the fixed point
$(f^*,G^*)$ of $\tilde{\cal R}$ with $G^*(x) = {1 \over 2}
[f^{*'}(x) - 1]$, whereas that near the right end is governded by the
fixed point with $G^*(x) = {1 \over 2} [f^{*'}(x) + 1]$.

Figure \ref{ASM2} shows the behavior of the asynchronous stability
multiplier $\lambda_{1,n}(A_\infty^{(3)},c)$ near the leftmost
critical line segment. The slopes $S_n$'s of $\lambda_{1,n}$ at both
ends obey well the one-term scaling law,
\begin{equation}
S_n \sim \nu^n \;\;{\rm for\; large\;} n,
\end{equation}
where $\nu = 3$. For any fixed value of $c$ inside the critical line
segment, $\lambda_{1,n}$ converges to zero as $n \rightarrow \infty$.
That is, all the interior points are critical points with
$\lambda_1^* = 0$. Hence the CB inside the critical line segment
becomes the same as that of the 1D map, which will be discussed in
more details below. This kind of 1D-like CB is governed by the fixed
point with $G^*(x) = {1 \over 2} f^{*'}(x)$, which has no relevant
CE's (see Table \ref{table1}).

For the case of a synchronous orbit in two linearly-coupled 1D
maps, its two Lyapunov exponents are given by
\begin{eqnarray}
\sigma_0(A) &=& \lim_{m \rightarrow \infty} {1 \over m}
\sum_{t=0}^{m-1} \ln|f'(x_t)|, \label{eq:1DLEXP} \\
\sigma_1(A,c) &=& \lim_{m \rightarrow \infty} {1 \over m}
\sum_{t=0}^{m-1} \ln|f'(x_t) - c|.
\label{eq:CLEXP}
\end{eqnarray}
Here $\sigma_0 (\sigma_1)$ is the synchronous (asynchronous) Lyapunov
exponents characterizing the mean exponential rate of divergence
of nearby orbits along (across) the symmetry line $y=x$. Note that the
synchronous Lyapunov exponent $\sigma_0$ is just that of the uncoupled
1D map and the coupling affects only the asynchronous Lyapunov
exponent $\sigma_1$. In order to see the phase dynamics near the
critical line segment in more details, we fix the value of the
nonlinearity parameter $A=A_\infty^{(3)}$ and obtain the
asynchronous Lyapunov exponent $\sigma_1$ of the synchronous orbit by
varying the coupling parameter $c$, which is shown in
Fig.~\ref{ALexp1}. (Note that the synchronous Lyapunov exponent
$\sigma_0$ of the synchronous quasiperiodic orbit becomes zero on the
$A=A_\infty^{(3)}$ line.) Inside the critical line segment
$(c_l < c < c_r)$, the synchronous quasiperiodic orbit on the $y=x$
symmetry line becomes a synchronous attractor with $\sigma_1 <0$.
Since the dynamics on the synchronous attractor is the same as that
for the uncoupled 1D case, the critical maps at interior points
exhibit essentially 1D-like CB. However, as the coupling parameter
$c$ passes through $c_l$ or $c_r$, the asynchronous Lyapunov exponent
$\sigma_1$ of the synchronous quasiperiodic orbit increases from zero,
and hence the coupling leads to desynchronization of the interacting
systems. Thus the synchronous quasiperiodic orbit ceases to be an
attractor outside the critical line segment, and new asynchronous
attractors appear. This is illustrated in Fig.~\ref{Desyn}.

We also study the critical scaling behavior of the asynchronous
Lyapunov exponent $\sigma_1$ near both ends for the case
$A=A_\infty^{(3)}$. As shown in Fig.~\ref{ALexp2}, the asynchronous
Lyapunov exponent $\sigma_1$ varies linearly with respect to $c$ near
both ends, i.e., $\sigma_1 \sim \varepsilon,\;\varepsilon
\equiv c - c^*$ $(c^*=c_l\;{\rm or}\;c_r)$. The critical scaling
behavior of $\sigma_1$ near both ends are obtained from the same CE
$\nu=3$ of the fixed points with $G^*(x) = {1 \over 2}
[1 \pm f^{*'}(x)]$. Consider a map with nonzero $\varepsilon$
(but with $A=A_\infty^{(3)})$ near both ends. It is then transformed
into a new one of the same form, but with a renormalized parameter
$\varepsilon'$ under a renormalization transformation. Here
the parameter $\varepsilon$ obeys a scaling law,
\begin{equation}
  \varepsilon' = \nu \, \varepsilon = 3 \, \varepsilon.
\end{equation}
Then the asynchronous Lyapunov exponent $\sigma_1$ satisfies the
homogeneity relation,
\begin{equation}
\sigma_1(\varepsilon') = 3\, \sigma_1(\varepsilon).
\end{equation}
This leads to the scaling relation,
\begin{equation}
\sigma_1(\varepsilon) \sim \varepsilon ^\eta,
\end{equation}
with exponent
\begin{equation}
\eta = \ln 3 / \ln \nu =1.
\end{equation}

We now study the CB of period quadruplings ($p=4$), as an example of
even period-$p$ tuplings. The stability diagrams of synchronous
orbits of period $q=4^n$ $(n=0,1,2,3)$ are shown in
Figs.~\ref{SDQ1}-\ref{SDQ2}.
The treelike structure of stability regions is similar to that for the
period-tripling case. As shown in Fig.~\ref{SDQ1}, four ``daughter''
quadrilaterals of level $1$ branch off from its ``mother''
quadrilateral of level $0$. However, an infinite number of successive
branchings occur only for the case of the central quadrilateral
including the $c=0$ line [see Fig.~\ref{SDQ2}(a)]. For the cases of
quadrilaterals other than the central one, successive
quadrilaterals of higher levels pile up without any branchings
[see Figs.~\ref{SDQ2}(b) and \ref{SDQ2}(c)].

Like the period-tripling case, there are two kinds of
period-quadrupling route which is a sequence of stability regions
with increasing period. The sequence of the quadrilaterals containing
the $c=0$ line segment, called the $Z_4$ route, converges to the
zero-coupling critical point
$(A_\infty^{(4)},0)$, where $A_\infty^{(4)}$
$(=1.942\,704\,354\,755\,467\,972\,167\,178 \dots)$ is the
accumulation point of the period-quadrupling sequence for the 1D case.
On the other hand, a sequence of quadrilaterals which piles up without
branchings, called an $L_4$ route, converges to a critical line.
For example, the leftmost one is the line joining two points
$(A_\infty^{(4)},c_l)$ and $(A_\infty^{(4)},c_r)$ [see
Fig.~\ref{SDQ2}(c)], where
$c_l=-3.888\,058\,931\,772\,634\,488 \dots$
and $c_r = -3.877\,063\,178\,096\,222\,051 \dots$~.
Note also that there are infinitely many $L_4$ routes.
Hence an infinite number of critical line segments, together with the
zero-coupling critical point, constitute the critical set.

We first consider the case of the $Z_4$ route ending at the
zero-coupling critical point. The critical scaling behavior for this
zero-coupling case is governed by the zero-coupling fixed map with
two relevant CE's $\nu_1 = \alpha$ $(=-38.819\,074 \dots)$ and
$\nu_2=4$.

Figure \ref{QPS} shows the quadrilaterals of the levels $n=1,2,3$ in
the $Z_4$ route. Note that Figs.~\ref{QPS}(a), \ref{QPS}(b), and
\ref{QPS}(c) nearly coincide. The height and width $h_n$ and $w_n$
\cite{note2} of the quadrilateral of level $n$ is therefore expected
to geometrically contract,
\begin{equation}
h_n \sim \delta^{-n},~ w_n \sim \alpha^{-n}~{\rm for ~ large~}n,
\end{equation}
where $\delta = 981.594\,976 \dots$~.
Thus the scaling factors of the nonlinearity and coupling parameters
become $\delta$ and $\alpha$, respectively, which will be shown
in more details below.

We follow, in the $Z_4$ route, the synchronous orbits of period
$q=4^n$ up to level $n=7$, and obtain a self-similar sequence of
parameters $(A_n,c_n)$, at which each orbit of level $n$ has some
given stability multipliers $(\lambda_0, \lambda_1)$.
Then the sequence $\{ (A_n,c_n) \}$ converges to the zero-coupling
critical point $(A_\infty^{(4)}, 0)$. Like the period-tripling case,
the two sequences
$\{ A_n \}$ and $\{ c_n \}$ obey well one-term scaling laws,
\begin{equation}
\Delta A_n \sim \delta^{-n},~ \Delta c_n \sim
\mu^{-n}~{\rm for~large~}n,
\end{equation}
where $\Delta A_n = A_n - A_{n-1}$ and $\Delta c_n = c_n - c_{n-1}$.
The scaling factors $\delta_n$ and $\mu_n$ of level $n$ converge to
their limit values $\delta$ $(=981.594\dots)$ and $\mu$
$(=-38.819 \dots)$, respectively, as shown in Table VI.
Note that $\delta$ is just the nonlinearity-parameter scaling factor
for the 1D case, and the value of the coupling-parameter scaling
factor $\mu$ agrees well with that of the first CE $\nu_1$
$(=\alpha)$. In fact, the second CE $\nu_2$ $(=4)$ also has an effect
on the scaling of the sequence $\{ \Delta c_n \}$. Thus it obeys well
the two-term scaling law,
\begin{equation}
\Delta c_n \sim C_1 \mu_1^{-n} + C_2 \mu_2^{-n} ~{\rm for~large~}n,
\end{equation}
where $|\mu_2| > |\mu_1|$, and $C_1$ and $C_2$ are some constants.
As shown in Table VII, the two scaling factors $\mu_1$ and $\mu_2$
are given by
\begin{equation}
\mu_1 = \nu_1,~ \mu_2= {\nu_1^2 \over \nu_2}.
\end{equation}

We also study the effect of CE's on the asynchronous stability
multipliers of synchronous orbits. Figure \ref{QASM1} shows three
plots of $\lambda_{1,n}(A_\infty^{(4)},c)$ versus $c$ for $n=1,2,3$.
The slope $S_n$ of $\lambda_{1,n}$ at the zero-coupling critical point
obeys well the two-term scaling law,
\begin{equation}
S_n \sim d_1 r_1^n + d_2 r_2^n ~{\rm for~large~}n,
\end{equation}
where $d_1$ and $d_2$ are some constants and $|r_1| > |r_2|$.
As shown in Table \ref{table8}, the scaling factors $r_{1,n}$ and
$r_{2,n}$ of level $n$ converge to constants $r_1$ $(=\nu_1)$ and
$r_2$ $(=\nu_2)$.

We next consider the cases of $L_4$ routes, each of which ends at a
critical line segment. As an example, consider the leftmost $L_4$
route [see Fig.~\ref{SDQ2}(c)], in which we follow two self-similar
sequences of parameters, one converging to the left end
$(A_\infty^{(4)},c_l)$ of the critical line segment and the other
one converging to the right end $(A_\infty^{(4)},c_r)$.
As in the case of the $Z_4$ route, the sequence $\{ A_n \}$ converges
geometrically to its limit value $A_\infty^{(4)}$ with the 1D scaling
factor $\delta$. The sequence $\{ c_n \}$ also obeys the one-term
scaling law,
\begin{equation}
\Delta c_n \sim \mu^{-n}~{\rm for~large~}n,
\end{equation}
where $\Delta c_n = c_n - c_{n-1}$.
The convergence of the scaling factor $\mu_n$ of level $n$ to its
limit value $\mu$ is shown in Table \ref{table9}. Note that the
scaling factors $\mu$ at both ends are the same, i.e. $\mu=4$.
Moreover, the critical asynchronous stability multipliers
$\lambda_1^*$'s at both ends are also the same, i.e. $\lambda_1^*=1$,
as shown in Table \ref{table9}. This is in contrast to the
period-tripling case where $\lambda_1^*$'s at both ends are different.
We also compare the values of $\mu$ and $\lambda_1^*$ with those of
the CE $\nu$ and $\lambda_1^*$ listed in Table \ref{table1} and find
that the CB at both ends is governed by the same fixed
point $(f^*,G^*)$ of $\tilde {\cal R}$ with $G^*(x) = {1 \over 2}
[f^{*'}(x) - 1]$, unlike the period-tripling case.

Figure \ref{QASM2} shows the behavior of the asynchronous stability
multiplier $\lambda_{1,n}(A_\infty^{(4)},c)$ near the leftmost
critical line segment. The growth of the slope $S_n$ of
$\lambda_{1,n}$ at both ends is governed
by the CE $\nu$ $(=4)$, i.e.,
\begin{equation}
S_n \sim \nu^n \;{\rm for \; large \;} n.
\end{equation}
However, for any fixed value of $c$ inside the critical line egment,
$\lambda_{1,n}$ converges to zero as $n \rightarrow \infty$.
Thus, all the interior points become critical ones with
$\lambda_1^*=0$. Consequently, the CB inside the critical
line segment becomes the same as that of the 1D map, as will be seen
below. This kind of 1D-like CB is governed by the fixed
point with $G^*(x) = {1 \over 2} f^{*'}(x)$, which has no relevant
CE's (see Table \ref{table1}).

In order to see the phase dynamics near the critical line segment in
more details, we fix the value of the nonlinearity parameter
$A=A_\infty^{(4)}$ and obtain the asynchronous Lyapunov exponent
$\sigma_1$ [see Eq.~(\ref{eq:CLEXP})] of the synchronous orbit by
varying the coupling parameter $c$, which is shown in
Fig.~\ref{QALexp1}. Inside the critical line segment ($c_l<c<c_r$),
the synchronous quasiperiodic orbit on the $y=x$ symmetry line becomes
a synchronous attractor with $\sigma_1 <0$, as shown in
Fig.~\ref{QDesyn}(a). Note that the dynamics on the
synchronous attractor
is the same as that for the uncoupled 1D case. Hence the critical maps
inside the critical line segment exhibit 1D-like CB.
However, as the coupling parameter $c$ passes through $c_l$
or $c_r$, the asynchronous Lyapunov exponent $\sigma_1$ of the
synchronous quasiperiodic orbit increases from zero, and hence the
coupling leads to desynchronization of the interacting systems. Thus
the synchronous quasiperiodic orbit ceases to be an attractor outside
the critical line segment, and new asynchronous attractors appear,
as shown in Figs.~\ref{QDesyn}(b) and \ref{QDesyn}(c).

We also study the scaling behavior of the asynchronous Lyapunov
exponent $\sigma_1$ near both ends of the critical line segment.
As shown in Fig.~\ref{QALexp2}, the asynchronous Lyapunov exponent
varies linearly with respect to $c$ near
both ends, i.e., $\sigma_1 \sim \epsilon$, $\epsilon \equiv c - c^*$
$(c^*=c_l$ or $c_r$). As in the period-tripling
case, using the scaling theory, one can also obtain the scaling
relation of $\sigma_1$, $\sigma_1 \sim \epsilon^\eta$
with exponent $\eta = \ln4 / \ln \nu =1$, where $\nu$ $(=4)$ is
the CE of the fixed point with $G^*(x) = {1 \over 2} [f^{*'}(x)-1]$.

Finally, we briefly summarize the results for the linear-coupling
case. The critical set consists of the zero-coupling critical point
and an infinite number of critical line segments. The CB at the
zero-coupling critical point is governed by the zero-coupling fixed
map with two CE's $\nu_1$ $(=\alpha)$ and $\nu_2$ $(=p)$ for all
period $p$-tupling cases. However, the CB near both ends of each
critical line segment depends on whether $p$ is even or odd. In the
case of odd $p$, the CB at one end is governed by a fixed point
$(f^*,G^*)$ of $\tilde {\cal R}$ with
$G^*(x) = {1 \over 2} [1 - f^{*'}(x)]$ and that at the other end
by another fixed point with $G^*(x) = {1 \over 2} [1 + f^{*'}(x)]$.
These two fixed points have only one CE $\nu=p$. On the other hand,
in the case of
even $p$, the CB at both ends is governed by the same fixed point
with $G^*(x) = {1 \over 2} [f^{*'}(x)-1]$.
Inside the critical line segment, the CB is the
same as that of the 1D map for all period $p$-tupling cases. This
kind of 1D-like CB is governed by the fixed point with
$G^*(x)={1 \over 2}f^{*'}(x)$, which has no relevant CE's.
Consequently, for even (odd) $p$, three (four) kinds of fixed
points govern the CB for the linearly-coupled case.

\subsection{Dissipatively coupled maps}
\label{sub:DCM}

As an example of the nonlinear-coupling case, we consider two
dissipatively-coupled 1D maps with the coupling function
\begin{equation}
g(x,y) = {c \over 2} [f(y) - f(x)],
\label{eq:DCFTN}
\end{equation}
and study the CB of the period triplings $(p=3)$ and
period quadruplings $(p=4)$.

Figures \ref{DCSD}(a) and \ref{DCSD}(b) show the stability diagrams of
synchronous orbits with period $q=p^n$ $(n=1,2,3)$ for $p=3$ and $4$,
respectively. As previously shown, each stability region of level $n$
(period $p^n$) in the parameter plane is bounded by four bifurcation
curves determined by $\lambda_{i,n} = \pm 1$ for $i=0,1$. An infinite
sequence of
such stability regions, called the ``period $p$-tupling route'',
converges to a critical line joining two ends $(A_\infty^{(p)},c^*_l)$
and $(A_\infty^{(p)},c^*_r)$, where $A_\infty^{(p)}$ is the
accumulation point of the period $p$-tupling sequence for the 1D case,
$c^*_l=0$ and $c^*_r=2$, as shown below. Hence only one critical line
segment constitutes the critical set for the dissipative case, unlike
the linearly-coupled case.

Consider two dissipatively-coupled 1D maps on the line
$A=A_\infty^{(p)}$ in the parameter plane, in which case the reduced
coupling function of the coupling function (\ref{eq:DCFTN}) is given
by
\begin{equation}
G(x) = {c \over 2} f_c'(x),
\end{equation}
where $f_c$ is the 1D critical map with the nonlinearity parameter
set to its critical value. By successive applications of the
renormalization operator $\tilde{\cal R}$ to $(f_c,G)$, we have
\begin{eqnarray}
f_n(x) &=& \alpha f_{n-1}^{(p)}({x \over \alpha}), \;\;
G_n(x) = {c_n \over 2} f_n'(x), \\
c_n &=& c_{n-1}^3 - 3 c_{n-1}^2 + 3 c_{n-1} \;{\rm for}\;p=3,
\label{eq:CRE3} \\
c_n &=& -c_{n-1}^4 + 4 c_{n-1}^3 - 6 c_{n-1}^2 + 4 c_{n-1}
\;{\rm for}\;p=4,
\label{eq:CRE4}
\end{eqnarray}
where $f_0 (x) = f_c (x)$, $G_0(x) = G(x)$, and $c_0=c$.
Here $f_n$ converges to the 1D fixed map $f^*(x)$.

The fixed points of the recurrence equations (\ref{eq:CRE3}) and
(\ref{eq:CRE4}) for $c$ are denoted by solid circles in
Fig.~\ref{DCFP}. For $p=3$, there are three fixed points,
\begin{equation}
c^* = 0,\, 1\, ,2,
\end{equation}
while only two fixed points,
\begin{equation}
c^* = 0,\,1\,,
\end{equation}
exist for $p=4$.
Stability of a fixed point $c^*$ is determined by its stability
multiplier $\lambda$ given by $\lambda = dc_n / dc_{n-1} |_{c^*}$.
Note that the fixed point at $c^*=1$ is superstable $(\lambda = 0)$.
The basin of attraction to the superstable fixed point becomes the
open interval $(0,2)$, because any initial $c$ inside the interval
$0 <c < 2$ converges to $c^*=1$. For the period-tripling case, both
ends $c^*_l=0$ and $c^*_r=2$ of the interval are unstable fixed points
with $\lambda=3$, and all points outside the interval diverge to the
plus or minus infinity. However, for the quadrupling case, only the
left end $c^*_l=0$, which is also the image of the right end $c^*_r=2$
under the recurrence relation (\ref{eq:CRE4}), is an unstable fixed
point with $\lambda =4$, and all points outside the interval diverge
to the minus infinity. Thus the line segment connecting two end points
$c^*_l=0$ and $c^*_r=2$ becomes the critical line for the
dissipative-coupling case.

All critical maps inside the critical line segment are attracted to
the fixed maps with the same reduced coupling function
$G^*(x) = {1 \over 2} f^{*'}(x)$. These fixed maps have no relevant
CE's, because the fixed point $c^*=1$ is superstable. Hence the
critical maps at interior points exhibit essentially 1D-like CB.
In the period-tripling case, the critical map at the left end $c^*_l$
is attracted to the zero-coupling fixed map,
\begin{equation}
{T^*}:
x_{t+1} =  {f^*}(x_t),\;\;
y_{t+1} = {f^*}(y_t),
\label{eq:ZCFM}
\end{equation}
with the reduced coupling function $G^*(x)=0$, and the critical map
at the right end $c_r^*$ to another fixed map,
\begin{equation}
{T^*}:
x_{t+1} =  {f^*}(y_t),\;\;
y_{t+1} = {f^*}(x_t),
\label{eq:CJFM}
\end{equation}
with the reduced coupling fixed function $G^*(x) = f^{*'}(x)$.
However, in the period-quadrupling case, the critical maps at both
ends are attracted to the zero-coupling fixed map (\ref{eq:ZCFM}).
Note that the two fixed maps of Eqs.~(\ref{eq:ZCFM}) and
(\ref{eq:CJFM}) have the same relevant CE's $\nu_1=\alpha$ and
$\nu_2=p$ (see Table \ref{table1}). However, for this
dissipatively-coupled case, the CB near both ends are governed only
by the second CE $\nu_2=p$ (i.e., the first relevant component
$\alpha_1$ in Eq.~(\ref{eq:Phin}) becomes zero), which can be easily
understood from the fact that the fixed points
$c^*=0,2$ are unstable ones with stability multiplier $\lambda=p$.

We study the critical scaling behavior associated with coupling
near the critical line segment and confirm the renormalization
results. For the dissipative-coupling case, the stability
multipliers of synchronous orbits with period $q=p^n$ $(p=3,4)$
become
\begin{equation}
\lambda_{0,n} (A) = \prod_{t=0}^{q-1} f'(x_t), \;\;
\lambda_{1,n} (A,c) = (1-c)^q \lambda_{0,n}.
\label{eq:DCSM}
\end{equation}
Let the pair of stability multipliers of the synchronous orbit
of level $n$ (period $p^n$) at a point $(A,c)$ be
$(\lambda_0,\lambda_1).$ Then there exists a ``conjugate point''
$(A,-c+2)$, at which the pair of stability multipliers becomes
$(\lambda_0,-\lambda_1)$ and $(\lambda_0, \lambda_1)$ for $p=3$ and
$4$, respectively. For $c=1$, $\lambda_1=0$ and the two conjugate
points become degenerate.

Like the linearly-coupled case, we follow the synchronous orbits of
level $n$ in the period $p$-tupling route, and obtain a self-similar
sequence of parameters $(A_n,c_n)$, at which each orbit of level $n$
has some given stability multipliers $(\lambda_0, \lambda_1)$.
Without loss of generality, we choose $\lambda_0=-1$. Then one can
find a pair of mutually conjugate sequences.
One sequence $\{ (A_n,c_n) \}$ can be obtained by fixing $-1 <
\lambda_1 <0$, which converges to the zero-coupling critical point
$(A_{\infty}^{(p)},0)$ as follows:
\begin{equation}
A_n - A_\infty^{(p)} \sim \delta^{-n}, \;\;
c_n = 1 - (-\lambda_1)^{p^{-n}} \sim - \ln(-\lambda_1) p^{-n}
\;\;{\rm for \; large \;}n,
\end{equation}
where $A_\infty^{(p)}$ and $\delta$ are just those for the 1D case.
Note that the coupling-parameter scaling factor is just the second
relevant CE $\nu_2$ $(=p)$ of the zero-coupling fixed map
(\ref{eq:ZCFM}). The ``conjugate sequence'' can also be obtained by
following $p^n$-periodic orbits with $\lambda_{1,n}= - \lambda_1$ and
$\lambda_1$ for $p=3$ and $4$, respectively. This kind of conjugate
sequence $\{ (A_n, c_n) \}$ converges to the right-end critical point
$(A_\infty^{(p)},2)$ (i.e., the conjugate point of the zero-coupling
critical point) as follows:
\begin{equation}
A_n - A_\infty^{(p)} \sim \delta^{-n}, \;\;
c_n -2 = -1 + (-\lambda_1)^{p^{-n}} \sim  \ln(-\lambda_1) p^{-n}
\;\;{\rm for \; large \;} n.
\end{equation}
Note also that the asymptotic scaling of the coupling-parameter
sequence is governed only by the second CE $\nu_2=p$ of the
``conjugate'' fixed map (\ref{eq:CJFM}) for $p=3$ and of the
zero-coupling fixed map (\ref{eq:ZCFM}) for $p=4$. Hence the critical
scaling behavior of the coupling parameter near both ends become the
same.

Figures \ref{DCSM}(a) and \ref{DCSM}(b) show three plots of
$\lambda_{1,n}(A_\infty^{(p)},c)$ versus $c$ for $n=1,2,3$ in the
period-tripling $(p=3)$ and period-quadrupling $(p=4)$ cases,
respectively. The critical asynchronous stability
multipliers $\lambda_1^*$'s of Eq.~(\ref{eq:CSM}) at both ends of the
critical line segment can be easily obtained from Eq.~(\ref{eq:DCSM}).
For $p=3$, $\lambda_1^* = \lambda^* (-\lambda^*)$ at the left (right)
end ($\lambda^*$ is the critical stability multiplier for the 1D
case), while for $p=4$, $\lambda_1^* = \lambda^*$ at both ends.
The slopes $S_n$'s of $\lambda_{1,n}$ at both ends also obey well the
same one-term scaling law,
\begin{equation}
S_n =
\left. {\displaystyle {{\partial \lambda_{1,n}} \over
{\partial c}} } \right|_{(A_\infty^{(p)},c^*)}
\sim  p^n \;\;{\rm for\;large\;} n,
\end{equation}
where $c^* = c^*_l$ or $c^*_r$.
Hence the growth of $S_n$ for large $n$ is governed only by the second
CE $\nu_2$ $(=p)$.

All interior points of the critical line segment become critical ones
with $\lambda_1^*=0$, because at any interior point $\lambda_{1,n}$
converges to zero as $n \rightarrow \infty$. Hence the CB inside the
critical line segment becomes the same as that of the 1D map, as will
be seen below. This kind of 1D-like CB is governed by the fixed maps
with the same reduced coupling function $G^{*'}(x) = {1 \over 2}
f^{*'}(x)$, which have no relevant CE's, as shown above.

For the dissipatively-coupled case, the asynchronous Lyapunov exponent
$\sigma_1$ is given by
\begin{equation}
\sigma_1(A,c) = \sigma_0(A) + \ln |1-c|,
\end{equation}
where
$\sigma_0$ is the 1D Lyapunov exponent of Eq.~(\ref{eq:1DLEXP}).
In order to see the phase dynamics near the critical line segment, we
fix the value of the nonlinearity parameter $A=A_\infty^{(p)}$. Then,
the asynchronous Lyapunov exponent becomes
$\sigma_1(A_\infty^{(p)},c) = \ln |1-c|$, because
$\sigma_0 (A_\infty^{(p)})=0$. Inside the critical line segment
($0<c<2$), the synchronous quasiperiodic orbit on the $y=x$ symmetry
line becomes a synchronous attractor with $\sigma_1 <0$ [see
Figs.~\ref{Desyn}(a) and \ref{QDesyn}(a)].
Note that the dynamics on the synchronous attractor is the same as
that for the uncoupled 1D case. Hence the critical maps inside the
critical line segment exhibit 1D-like CB.
However, as the coupling parameter $c$ passes through $c^*_l$ or
$c^*_r$, the asynchronous Lyapunov exponent $\sigma_1$ of the
synchronous quasiperiodic orbit increases from zero, and hence the
coupling leads to desynchronization of the interacting systems. Thus
the synchronous quasiperiodic orbit ceases to be an attractor outside
the critical line segment, and new asynchronous attractors appear,
as shown in Fig.~\ref{DCDesyn}.

We also study the scaling behavior of the asynchronous Lyapunov
exponent $\sigma_1$ near both ends of the critical line segment.
The asynchronous Lyapunov exponent near both ends becomes $\sigma_1
\sim \epsilon$, where $\epsilon \equiv c - c^*$ $(c^*=c_l^*$ or
$c_r^*$). As in the linearly-coupled case, using the scaling theory,
one can also obtain the scaling relation of $\sigma_1$, $\sigma_1
\sim \epsilon^\eta$ with exponent $\eta = \ln p / \ln \nu_2 =1$,
where $\nu_2$ $(=p)$ is the second CE of the fixed maps
(\ref{eq:ZCFM}) and (\ref{eq:CJFM}).

\section{Extension to many coupled maps}
\label{sec:EXMCM}

In this section we study the CB of period $p$-tuplings in $N$
$(N \geq 3)$ coupled 1D maps, in which the coupling extends to the
$K$th $[1 \leq K \leq {N \over 2} ({{N-1} \over 2})$ for even (odd)
$N$] neighbor(s) with equal strength. It is found that the CB depends
on the range of coupling. In the global-coupling case, in which each
1D map is coupled to all the other 1D maps with equal strength, both
the structure of the critical set and the CB are the same as those
for the two-coupled case, irrespectively of $N$.
However, for the cases of nonglobal couplings of shorter range,
a significant change in the structure of the critical set may
or may not occur according as the coupling is linear or not.
As examples of the linear and nonlinear nonglobal couplings, we study
the linearly- and diffusively-coupled, nearest-neighbor coupling
cases, respectively. For the linearly-coupled case, of the infinite
number of period $p$-tupling routes for the global-coupling case,
only the route ending at the zero-coupling critical point is left in
the parameter plane. On the other hand, for the diffusively-coupled
case, one critical line segment constitutes the critical set, as in
the globally-coupled case.

\subsection{Stability of periodic orbits in many coupled maps}
\label{sub:MCM}

Consider $N$ symmetrically coupled 1D maps with a periodic
boundary condition,
\begin{eqnarray}
T: x_m(t+1) &=& F(\sigma^{m-1} {\bf x}(t) ) \nonumber \\
  &=& F(x_m(t), x_{m+1}(t), \dots , x_{m-1}(t)), \nonumber \\
  && \;\;\;\;\;\;\;\;\;\;\;\;\;\;\;\;\;\;\;\;\;\;\;\;\;\;\;\;\;
   m=1, \dots, N,
\label{eq:MCM}
\end{eqnarray}
where $N$ is a positive integer larger than or equal to $2$,
${\bf x} = (x_1, \dots , x_N)$ and $\sigma$ is the cyclic
permutation of $\bf x$ [i.e.,
$\sigma {\bf x} =(x_2, \dots , x_1)$].
Here $x_m(t)$ is the state of the $m$th element at a discrete time
$t$, and the periodic condition imposes $x_m(t) = x_{m+N}(t)$ for all
$m$. Like the two-coupled case with $N=2$, the function $F$ consists
of two parts:
\begin{equation}
 F({\bf x}) = f(x_1) + g ({\bf x}),
\end{equation}
where $f$ is an uncoupled 1D map with a quadratic maximum at $x=0$,
and $g$ is a coupling function. The unoupled 1D map $f$ satisfies
the normalization condition (\ref{eq:NC}),
and the coupling function $g$ obeys the condition
\begin{equation}
g(x,\dots,x)=0 \;{\rm for}\;{\rm any}\;x.
\label{eq:MCCC}
\end{equation}

The $N$-coupled map $T$  has a cyclic permutation symmetry,
\begin{equation}
\sigma^{-1} T \sigma ({\bf x}) = T ({\bf x}) \;{\rm for}\;
{\rm all}\;{\bf x},
\label{eq:CPS}
\end{equation}
where $\sigma^{-1}$ is the inverse of $\sigma$.
The set of all fixed points of $\sigma$ forms a symmetry line
on which
\begin{equation}
x_1 = \cdots = x_N.
\label{eq:MCSL}
\end{equation}
It follow from Eq.~(\ref{eq:CPS}) that the cyclic permutation $\sigma$
commutes with the map $T$, i.e., $\sigma T = T \sigma$.
Hence the symmetry line becomes invariant under $T$, i.e., if a point
${\bf x}$ lies on the symmetry line, then its image $T({\bf x})$ also
lies on it. An orbit is called a synchronous orbit if it lies on
the symmetry line, i.e., it satisfies
\begin{equation}
x_1(t) = \cdots = x_N(t) \equiv x(t) \;{\rm for\;all}\;t.
\label{eq:MCSO}
\end{equation}
Otherwise, it is called an asynchronous orbit. Here we study only the
synchronous orbits. They can be easily found from the uncoupled 1D
map, $x(t+1) = f(x(t))$, because of the condition (\ref{eq:MCCC}).

Consider an element, say the $m$th element, in the $N$-coupled map
$T$. Then the $(m \pm \delta)$th elements are called the
$\delta$th neighbors of the $m$th element, where
$1 \leq \delta \leq {N \over 2} ({{N-1} \over 2})$ for even (odd)
$N$. If the coupling extends to the $K$th neighbor(s), then the number
$K$ is called the range of coupling.

A general form of coupling for odd $N$ $(N \geq 3)$ is given by
\begin{eqnarray}
g(x_1,\dots,x_N) &=& {\frac {c} {2K+1}}
  {\sum_{m=-K}^{K}} [u(x_{1+m}) - u(x_1)], \nonumber \\
&=& c \left[ {\frac {1}{2K+1}} {\sum_{m=-K}^{K}} u(x_{1+m}) - u(x_1)
\right], \nonumber \\
&&K=1,\dots,{\frac {N-1} {2}},
\label{eq:MCCF}
\end{eqnarray}
where $c$ is a coupling parameter and $u$ is a function of one
variable. Here the coupling extends to the $K$th neighbors with
equal coupling strength, and the function $g$ satisfies the condition
(\ref{eq:MCCC}).
The extreme long-range interaction for
$K= {{\frac {N-1} {2}}}$ is called a global coupling, for  which  the
coupling  function $g$ becomes
\begin{eqnarray}
g(x_1,\dots,x_N)  &=&  {\frac  {c}  {N}}{\sum_{m=1}^{N}}  [u(x_{m}) -
u(x_1)] \nonumber \\
&=& c  \left[ {\frac  {1}{N}} {\sum_{m=1}^{N}} u(x_m)  - u(x_1)
\right].
\label{eq:GC}
\end{eqnarray}
This  is a  kind  of mean-field  coupling,  in which  each  element is 
coupled to all the  other elements  with equal  coupling strength. All
the other couplings with $K < {{\frac {N-1} {2}}}$ (e.g.,
nearest-neighbor coupling with $K=1$) will be referred to as
non-global couplings. The $K=1$ case for $N=3$ corresponds to both the
global coupling and the nearest-neighbor coupling.

We next consider the case of even $N$ $(N \geq 2)$. The  form of
coupling of  Eq.~(\ref{eq:MCCF}) holds  for the  cases of
non-global couplings with $K=1,\dots,{{\frac{N-2}{2}}}$ $(N \geq 4)$.
The global coupling for $K= {{\frac {N} {2}}}$ $(N \geq 2)$ also
has the form of Eq.~(\ref{eq:GC}), but it cannot  have the form of
Eq.~(\ref{eq:MCCF}), because there exists only one farthest neighbor
for $K= { {\frac{N}{2}} }$, unlike the case of odd $N$.
The  $K=1$ case  for $N=2$  also  corresponds to  the nearest-neighbor 
coupling as well as to the global coupling, like the $N=3$ case.

The stability analysis of an orbit in many coupled maps is
conveniently carried out by Fourier-transforming with respect to the
discrete space $\{m\}$ \cite{Kapral}. Consider an orbit
$\{ {x_m}(t)\;  ; \;m=1,\dots,N \}$ of the $N$ coupled maps
(\ref{eq:MCM}).
The discrete spatial Fourier transform of the orbit is:
\begin{eqnarray}
{\cal F}[{x_m(t)}] &\equiv& {\frac{1}{N}} {\sum_{m=1}^{N}}
{e^{-2{\pi}imj/N}} {x_m}(t) = {\xi}_j(t), \nonumber \\
&&\;\;\;\;\;\;\;\;\;\; j=0,1,\dots,N-1.
\label{eq:FT}
\end{eqnarray}
The Fourier transform $\xi_j(t)$ satisfies $\xi_j^*(t) = \xi_{N-j}(t)$
($*$ denotes complex conjugate), and
the wavelength of a mode with index $j$ is ${\frac {N}{j}}$ for
$j \leq {{ {\frac {N} {2}}}}$ and
${\frac {N} {N-j}}$ for $j > {{\frac {N} {2}}}$.

To  determine the  stability of  a synchronous orbit
[$x_1(t)  = \cdots =x_N(t) \equiv x(t)$ for all $t$],
we consider an infinitesimal perturbation $\{ {\delta}x_m(t) \}$
to the synchronous orbit, i.e.,
$x_m(t)=x(t)+{\delta}x_m(t)$ for $m=1,\dots,N$.
Linearizing the $N$-coupled map (\ref{eq:MCM}) at the
synchronous orbit, we obtain:
\begin{equation}
{\delta}{x_m} (t+1) = f'(x(t)) {\delta x_m}(t) +
 {\sum_{l=1}^{N}} {G^{(l)}}(x(t))\;{\delta} x_{l+m-1},
\label{eq:MCLE}
\end{equation}
where
\begin{equation}
G^{(l)}(x) \equiv
\left. { \frac{\partial g(x_1,\dots,x_N)}{\partial x_l} }
\right |_{x_1=\cdots=x_N=x}.
\label{eq:RCF}
\end{equation}
Hereafter the  functions $G^{(l)}$'s  will be called
``reduced'' coupling functions
of $g(x_1,\dots,x_N)$.

Let  ${\delta  {\xi}_j}(t)$  be  the  Fourier  transform  of
{$\delta x_m(t)$}, i.e.,
\begin{eqnarray}
\delta \xi_j =
{\cal F}[{\delta x_m(t)}] &=& {\frac{1}{N}} {\sum_{m=1}^{N}}
{e^{-2{\pi}imj/N}} {\delta x_m}, \nonumber \\
&&\;\;\;\;\;\;\;\;\;j=0,1,\dots,N-1.
\end{eqnarray}
Here $\delta \xi_0$ is the synchronous-mode perturbation, and all
the other $\delta \xi_j$'s with nonzero indices $j$ are the
asynchronous-mode perturbations. Then the Fourier transform of
Eq.~(\ref{eq:MCLE}) becomes:
\begin{eqnarray}
\delta {\xi}_j (t+1) &=& [f'(x(t)) \nonumber \\
&& + \sum_{l=1}^{N} {G^{(l)}}(x(t)) {e^{2 \pi i(l-1)j/N}}]
{\delta {\xi}_j}(t), \nonumber \\
&&\;\;\;\;\;\;\;\;\;\;\;\;\;\;\; j=0,1,\dots,N-1.
\label{eq:MCLM1}
\end{eqnarray}
Note that all the modes $\delta \xi_j$'s become decoupled for the
synchronous orbit.

For a synchronous orbit with period $q$, its linear stability is
determined by iterating the linearized map (\ref{eq:MCLM1}) $q$
times:
\begin{eqnarray}
\delta {\xi}_j (t+1) &=& \prod_{m=0}^{q-1} [f'(x(t+m)) \nonumber \\
&& + \sum_{l=1}^{N} {G^{(l)}}(x(t+m)) {e^{2 \pi i(l-1)j/N}}]
\delta {\xi}_j (t)
\nonumber \\
&&\;\;\;\;\;\;\;\;\;\;\;\;\;\;\;\;\;\;\;\; j=0,1,\dots,N-1.
\label{eq:MCLM2}
\end{eqnarray}
That is, the stability multipliers of the orbit are given by
\begin{eqnarray}
\lambda_j &=& \prod_{t=0}^{q-1} [f'(x(t))
+ \sum_{l=1}^{N} {G^{(l)}}(x(t)) {e^{2 \pi i(l-1)j/N}}],
\nonumber \\
&&\;\;\;\;\;\;\;\;\;\;\;\;\;\;\; j=0,1,\dots,N-1.
\label{eq:MCSM}
\end{eqnarray}
Here the first stability multiplier $\lambda_0$ is associated with
the stability against the synchronous-mode perturbation, and hence
it may be called the synchronous stability multiplier. On the
other hand, all the other stability multipliers $\lambda_j$'s
$(j \neq 0)$ are called the asynchronous stability multipliers,
because they are associated with the stability against the
asynchronous-mode perturbations.

A synchronous orbit becomes stable when it is stable against all
the synchronous-mode and asynchronous-mode perturbations, i.e.,
the moduli of all stability multipliers are less than unity
$(|\lambda_j| <1$ for $j=0,\dots,N-1$).
Hence the stable region of the synchronous orbit in the
parameter plane is bounded by the synchronous and asynchronous
bifurcation lines determined by the equations $\lambda_j = \pm 1$
for $j=0,\dots,N-1$. When the $\lambda_0 = 1 (-1)$ line is
crossed, the synchronous orbit loses its stability via synchronous
saddle-node (period-doubling) bifurcation. However,
when the $\lambda_j$ ($j \neq 0) = 1 (-1)$ line is crossed,
it becomes unstable via asynchronous pitchfork (period-doubling)
bifurcation.

It follows from Eq.~(\ref{eq:MCCC}) that
\begin{equation}
\sum_{l=1}^N G^{(l)}(x) =0.
\label{eq:SR}
\end{equation}
Hence the synchronous stability multiplier $\lambda_0$ for $j=0$
becomes
\begin{equation}
\lambda_0 = \prod_{t=0}^{q-1} f'(x(t)),
\label{eq:MCSSM}
\end{equation}
which is just the stability multiplier of the uncoupled 1D map.
While there is no coupling effect on $\lambda_0$, the coupling
generally affects asynchronous stability multipliers
$\lambda_j$'s of $j \neq 0$. The effect of the coupling on the
asynchronous stability multipliers depends on the range of coupling,
as will be seen in the next two subsections.

\subsection{Global-coupling case}
\label{sub:GC}

In this subsection, we study the CB of period $p$-tuplings in many
coupled maps with a global coupling. It is shown that both the
structure of the critical set and the CB for the case of $N$
globally-coupled maps are the same as those for the case of two
coupled maps, independently of $N$.

In case of the global coupling (\ref{eq:GC}), the reduced coupling
functions of Eq.~(\ref{eq:RCF}) become:
\begin{equation}
{G^{(l)}}(x) = \left \{
 \begin{array}{l}
  (1-N) G(x)\;\;\;\; {\rm for}\;l=1, \\
  \;\;\;\;\;\;G(x)\;\;\;\;\;\;\;\;\;\;{\rm for}\;l \neq 1,
  \end{array}
  \right.
\label{eq:GCR}
\end{equation}
where $G(x)= {{\frac{c}{N}}} u'(x)$.
Substituting $G^{(l)}$'s into Eq.~(\ref{eq:MCSM}), we find that
all the asynchronous stability multipliers are the same:
\begin{eqnarray}
\lambda_1 = \cdots = \lambda_{N-1} &=&
\prod_{t=0}^{q-1} [f'(x(t)) - N G(x(t))] \nonumber \\
&=& \prod_{t=0}^{q-1} [f'(x(t)) - c u'(x(t))].
\label{eq:GCASSM}
\end{eqnarray}
Hence there exist only two independent stability multipliers
$\lambda_0$ and $\lambda_1$ $(=\lambda_1 = \cdots = \lambda_{N-1}$)
for the global-coupling case. Note also that the values of
$\lambda_0$ and $\lambda_1$ are independent of $N$ and they are
the same as those of two coupled maps. Thus the stability diagram
of synchronous orbits of period $p^n$ $(n=0,1,2,\dots)$ in any $N$
globally coupled maps becomes the same as that of the two coupled
maps. Consequently, the two parameter scaling factors associated
with scaling of the nonlinearity and coupling parameters are the
same as those of the two coupled maps, independently of $N$.
That is, the CB of $N$ globally coupled maps becomes the same as that
for the case of the two coupled maps, irrespectively of $N$, which
is also shown below by a renormalization analysis.

We now follow the same procedure of Sec.~\ref{sub:RA} and
straightforwardly extend the renormalization results of the two
coupled maps to many globally
coupled maps. The rescaling operator $B$ of Eq.~(\ref{eq:SO})
becomes $\alpha I$ for the case of $N$ coupled maps, where $I$ is the
$N \times N$ identity matrix. Applying the period $p$-tupling
operator $\cal N$ of Eq.~(\ref{eq:RON}) to the $N$-coupled maps
(\ref{eq:MCM}) $n$ times, we obtain the $n$-times renormalized map
$T_n$ of the form,
\begin{eqnarray}
{T_n}: x_m(t+1) &=& F_n (x_m(t),x_{m+1}(t),\dots,x_{m-1}(t))
                   \nonumber \\
       &=& f_n(x_m(t)) + g_n(x_m(t),x_{m+1}(t),\dots,x_{m-1}(t)),
       \nonumber \\
  && \;\;\;\;\;\;\;\;\;\;\;\;\;\;\;\;\;\;\;\;\;\;\;\;\;\;\;\;\;
   m=1, \dots, N.
\label{eq:MCRTn}
\end{eqnarray}
Here $f_n$ and $g_n$ are the uncoupled and coupling parts of the
$n$-times renormalized function $F_n$, respectively.
They satisfy the following recurrence equations:
\begin{eqnarray}
&f_{n+1}(x_1) = & \alpha f_n^{(p)} ({\frac {x_1} {\alpha}}),
\label{eq:MCRUCFn} \\
&g_{n+1}({\bf x}) =& \alpha F_n^{(p)} ({\frac {{\bf x}} {\alpha}})
-{\alpha} f_n^{(p)} ({\frac {x_1} {\alpha}}),
\label{eq:MCRCFn}
\end{eqnarray}
where
\begin{eqnarray}
&f_n^{(p)} (x_1)& = f_n (f_n^{(p-1)}(x_1)), \\
&F_n^{(p)}({\bf x})& = F_n(F_n^{(p-1)}({\bf x}),
F_n^{(p-1)}(\sigma {\bf x}), \dots, F_n^{(p-1)}(\sigma^{N-1}
 {\bf x})),
\end{eqnarray}
and the rescaling factor is chosen
to preserve the normalization condition $f_{n+1}(0)=1$, i.e.,
$\alpha = {1 \over f^{(p-1)}_n(1)}$.
Then, the recurrence relations (\ref{eq:MCRUCFn}) and
(\ref{eq:MCRCFn}) define a renormalization operator ${\cal R}$ of
transforming a pair of functions $(f,g)$:
\begin{equation}
\left( 
\begin{array}{c}
{f_{n+1}} \\ 
{g_{n+1}}
\end{array}
\right) = {\cal R} \left( 
\begin{array}{c}
{f_n} \\ 
{g_n}
\end{array}
\right).  \label{eq:MCRORn}
\end{equation}

A critical map with the nonlinearity and coupling parameters set to
their critical values is attracted to a fixed map $T^*$ under the
iterations of the renormalization transformation $\cal N$:
\begin{eqnarray}
{T^*}: x_m(t+1) &=& F^* (x_m(t),x_{m+1}(t),\dots,x_{m-1}(t))
                   \nonumber \\
       &=& f^*(x_m(t)) + g^*(x_m(t),x_{m+1}(t),\dots,x_{m-1}(t)),
       \nonumber \\
  && \;\;\;\;\;\;\;\;\;\;\;\;\;\;\;\;\;\;\;\;\;\;\;\;\;\;\;\;\;
   m=1, \dots, N,
\label{eq:MCFM}
\end{eqnarray}
where $(f^*,g^*)$ is the fixed point of $\cal R$ with
$\alpha = 1/f^*(1)$. Since $f^*$ is just the 1D fixed map, only the
equation for the coupling fixed function $g^*$ is left to be solved.

As in the two-coupled case, we construct a tractable recurrence
equation for the reduced coupling function $G^{(l)}(x)$.
That is, differenting the recurrence equation
(\ref{eq:MCRORn}) with respect to $x_l$ $(l=2,\dots,N)$ and setting
$x_1 = \cdots = x_N = x$, we obtain \cite{note3}
\begin{eqnarray}
G^{(l)}_{n+1}(x) &=& F_{n,l}^{(p)}({x \over \alpha}) \nonumber \\
&=& F_{n,l}^{(p-1)} ({x \over \alpha})
{f'_n}({f_n^{(p-1)}}({x \over \alpha})) +
f^{(p-1)'}_n({x \over \alpha}) G^{(l)}_{n}({f_n^{(p-1)}}
({x \over \alpha})) \nonumber \\
&& + \sum_{m=1}^N F_{n,l-m+1}^{(p-1)} ({x \over \alpha})
[G^{(m)}_n (f_n^{(p-1)}({ x \over \alpha})) -
 G^{(l)}_n (f_n^{(p-1)}({ x \over \alpha}))], \nonumber \\
 &&\;\;\;\;\;\;\;\;\;\;\;\;\;\;\;\;\;\;\;\;\;\;\;\;\;
 l=2,\dots,N.
\label{eq:MCRRE}
\end{eqnarray}
where
$F^{(p)}_{n,m}(x) \equiv \partial F^{(p)}_n ({\bf x}) /
\partial x_m|_{x_1 = \cdots =x_N =x}$ $(m=1,\dots,N)$.
Note that these reduced coupling functions satisfy the sum rule
of Eq.~(\ref{eq:SR}) (i.e., $\displaystyle{ \sum_{l=1}^N
G^{(l)}_n(x) =0}$), and $G^{(l)}_n(x) = G^{(l+N)}_n(x)$
[or equivalently, $F^{(p)}_{n,m}(x) = F^{(p)}_{n,m+N}(x)$]
due to the periodic condition.

In the global-coupling case, the initial reduced coupling functions
$\{ G^{(l)}(x) \}$ satisfy Eq.~(\ref{eq:GCR}), i.e., there
exists only one independent reduced coupling function $G(x)$.
Then, it is easy to see that the successive images
$\{ G^{(l)}_n (x) \}$ of $\{ G^{(l)}(x) \}$ under the transformation
(\ref{eq:MCRRE}) also satisfy Eq.~(\ref{eq:GCR}), i.e.,
\begin{eqnarray}
&& G^{(2)}_n (x) = \cdots = G^{(N)}_n (x) \equiv G_n(x),  \\
&& [{\rm or\;equivalently},
 F^{(p)}_{n,2}(x) = \cdots = F^{(p)}_{n,N}(x)
 \equiv F^{(p)}_n(x).]
\end{eqnarray}
Consequently, there remains only one recurrence equation for the
independent reduced coupling function $G(x)$:
\begin{eqnarray}
G_{n+1}(x) &=& F_n^{(p)}({x \over \alpha}) \nonumber \\
&=& F_n^{(p-1)} ({x \over \alpha})
[{f'_n}({f_n^{(p-1)}}({x \over \alpha}))-
NG_{n}({f_n^{(p-1)}}({x \over \alpha}))] \nonumber \\
&& +
f^{(p-1)'}_n({x \over \alpha}) G_{n}({f_n^{(p-1)}}({x \over \alpha})).
\label{eq:MCIRRE}
\end{eqnarray}
Then, together with Eq.~(\ref{eq:MCRUCFn}), Eq.~(\ref{eq:MCIRRE})
defines a reduced renormalization operator $\tilde{\cal R}$
of transforming a pair of functions $(f,G)$ such that
$(f_{n+1},G_{n+1}) = {\tilde{\cal R}} (f_n,G_n)$.

We look for fixed points $(f^*, G^*)$ of $\tilde {\cal R}$ such that
$(f^*,G^*)={\tilde {\cal R}} (f^*,G^*)$. Here $f^*(x)$ is just the
1D fixed function. Only the equation for $G^*$ is therefore left to be
solved. Since the transformation (\ref{eq:MCIRRE}) for $G$ holds for
any globally-coupled-map cases with $N \geq 2$, it can be regarded as
a generalized version of Eq.~(\ref{eq:RRE}) for the two-coupled case.
Comparing the expression in Eq.~(\ref{eq:MCIRRE})
with that in Eq.~(\ref{eq:RRE}), one can easily see that they are the
same except for the factor $N$. Making a change of the independent
reduced coupling function $G(x) \rightarrow {2 \over N} G(x)$
[equivalently, $F^{(p)}(x) \rightarrow {2 \over N} F^{(p)}(x)$],
Eq.~(\ref{eq:MCIRRE}) is transformed into Eq.~(\ref{eq:RRE}).
Consequently, rescaling the solutions (\ref{eq:RCFFTN}) for the
two-coupled case with the scaling factor $2 \over N$, one can obtain
the solutions for the case of $N$ globally coupled maps:
\begin{mathletters}
\label{eq:MCRCFFTN}
\begin{eqnarray}
G^*(x) &=& 0~~~{\rm for~all~}p, \\
G^*(x) &=& {1 \over N} f^{*'}(x)~~~{\rm for~all~}p, \\
G^*(x) &=& {1 \over N} [f^{*'}(x) - 1]~~~{\rm for~all~}p, \\
G^*(x) &=& {1 \over N} [f^{*'}(x) + 1]~~~{\rm for~odd~}p, \\
G^*(x) &=& {2 \over N} f^{*'}(x) ~~~{\rm for~odd~}p.
\end{eqnarray}
\end{mathletters}
\noindent
Thus, there exist three (five) fixed points $(f^*,G^*)$ of
$\tilde{\cal R}$
for the case of even (odd) p, independently of $N$.

For the same reason as for the two coupled maps [see
Eq.~(\ref{eq:1CSM})], the critical stability multipliers have the
values of the stability multipliers of the fixed point of the fixed
map $T^*$. From Eqs.~(\ref{eq:MCSSM}) and (\ref{eq:GCASSM}), we
obtain two independent critical stability multipliers
$\lambda^*_0$ and $\lambda^*_1$:
\begin{equation}
\lambda_0^* =  f^{*'}({\hat x}), \;\;\;
\lambda_1^* = f^{*'}({\hat x})-NG^*({\hat x}),
\label{eq:MC1CSM}
\end{equation}
where $\hat x$ is the fixed point of the 1D fixed function [i.e.,
${\hat x}= f^* ({\hat x})$] and $\lambda^*_0$ is just the critical
stability multiplier $\lambda^*$ of the uncoupled 1D map.
Substituting $G^*$'s into Eq.~(\ref{eq:MC1CSM}), we obtain
the same critical asynchronous stability multipliers $\lambda^*_1$'s
as in the case of two coupled maps, as listed in Table \ref{table10}.

Consider an infinitesimal perutrbation $(h,\Phi)$ to a fixed point
$(f^*,G^*)$ of the reduced renormalization operator $\tilde{\cal R}$.
Linearizing $\tilde{\cal R}$ at the fixed point, we obtain the
recurrence
equation for the evolution of $(h,\Phi)$, $(h_{n+1},\Phi_{n+1})
= {\tilde {\cal L}} (h_n,\Phi_n)$. As in the two-coupled case,
the linearized operator $\tilde{\cal L}$ also has a semi-block form
[i.e., $h_{n+1}(x) = [{\tilde{\cal L}}_1 h_n] (x)$ and $\Phi_{n+1}(x)
= [{\tilde{\cal L}}_2 \Phi_n](x) + [{\tilde{\cal L}}_3 h_n](x)]$.
It follows from the reducibility of $\tilde{\cal L}$
into a semi-block form that one can find eigenvalues of ${\tilde{\cal
L}}_1$ and ${\tilde{\cal L}}_2$ separately and then they give the
whole spectrum of $\tilde{\cal L}$.

All the fixed points $(f^*,G^*)$
have a common relevant eigenvalue $\delta$ of ${\tilde{\cal L}}_1$
(i.e., the relevant eigenvalue for the case of the uncoupled
1D maps) associated with the critical scaling of the nonlinearity
parameter of the uncoupled 1D map. However, the relevant CE's of
${\tilde{\cal L}}_2$, associated with the critical scaling of the
coupling parameter, depend on the kind of the fixed points, as
in the case of the two coupled maps. Consider an infinitesimal
perturbation $\Phi$ to a fixed point $G^*$ of the
recurrence equation (\ref{eq:MCIRRE}). Linearizing
Eq.~(\ref{eq:MCIRRE}) at the fixed point $G^*$, we obtain an
equation for the evolution of $\Phi$
[i.e., $\Phi_{n+1} (x) = {\tilde{\cal L}}_2 \Phi_n(x)$].
Note again that Eq.~(\ref{eq:MCIRRE}) is transformed into the
recurrence equation (\ref{eq:RRE}) for the two-coupled case under a
mere scale change of the independent reduced coupling function
$G(x) \rightarrow {2 \over N} G(x)$. Consequently, the CE equation
[i.e., ${\tilde{\cal L}}_2 \Phi^*(x) = \nu \Phi^*(x)$] for the
case of $N$ globally coupled maps becomes the same as that for the
case of two coupled maps, independently of $N$. Then, following the
same procedure of Sec.~{\ref{sub:RA}, one can obtain the same
CE's as those for the case of two coupled maps, as listed in
Table \ref{table10}.

\subsection{Nonglobal-coupling cases}
\label{sub:NGC}

In this subsection, we choose $f(x) = 1 - A x^2$ as the uncoupled
1D map and study the CB of the period $p$-tuplings
for the nonglobal-coupling cases. A significant change in the
structure of the critical set may or may not occur according as the
coupling is linear or not. As examples of the linear and nonlinear
nonglobal couplings, we study the linearly- and diffusively-coupled,
nearest-neighbor coupling cases, respectively.
For the linearly-coupled case, only the zero-coupling critical point
is left in the parameter plane, which is in contrast to the
global-coupling case. On the other hand, for the diffusively-coupled
case, one critical line segment constitutes the critical set, as in
the global-coupling case.

Consider a nonglobal coupling of the form (\ref{eq:MCCF}) and define
\begin{equation}
G(x) \equiv {c \over {2K+1}} u'(x),
\end{equation}
where $1 \leq K \leq {{N-2} \over 2} ({{N-3} \over 2})$ for even (odd)
$N$ larger than $3$. Then we have
\begin{equation}
{G^{(l)}}(x) = \left \{
 \begin{array}{l}
  -2 K G(x)\;\;\; {\rm for}\;l=1, \\
  \;\;\;\;\;G(x)\;\;\;\;\;\; {\rm for}\;  2 \leq l \leq  1+K \;\; {\rm 
or} \\
  \;\;\;\;\;\;\;\;\;\;\;\;\;\;\;\;\;\;\;{\rm  for}\;N+1-K \leq  l \leq 
N, \\
  \;\;\;\;\;\;\;0\;\;\;\;\;\;\;\;\;\;{\rm otherwise.}
  \end{array}
  \right.
\end{equation}
Substituting $G^{(l)}$'s into Eq.~(\ref{eq:MCSM}), we find that
each stability multiplier $\lambda_j$ (associated with
the stability against the $j$th-mode perturbation) is given by
\begin{equation}
\lambda_j =  \prod_{t=0}^{q-1} [f'(x(t)) - S_N(K,j) c u'(x(t))],
\label{eq:NGCASSM}
\end{equation}
where
\begin{equation}
{S_N}(K,j) \equiv {4 \over {2K+1}} {\sum_{k=1}^{K}}
\sin^2 {{\pi jk} \over {N}}
= 1- {\frac {\sin(2K+1) {{\frac{\pi j}{N}}}}
{(2K+1) \sin{{\frac{\pi j}{N}}}}}.
\label{eq:SF}
\end{equation}
Hence, unlike the global-coupling case [see Eq.~(\ref{eq:GCASSM})],
the stability multipliers vary depending on the
coupling range $K$ as well as on  the mode number $j$.
Since $S_N(K,j)  =  S_N(K,N-j)$, the stability multipliers satisfy
\begin{equation}
\lambda_j = \lambda_{N-j},\;\;j=0,1,\dots,N-1.
\end{equation}
Thus it is sufficient to consider only the case of
$0 \leq j \leq {N \over 2}$ $({{N-1} \over 2})$ for even (odd) $N$.
Comparing the expression in Eq.~(\ref{eq:NGCASSM}) with that in
Eq.~(\ref{eq:GCASSM}) for $j \neq 0$, one can easily see that
they are the same except for the factor $S_N (K,j)$. Consequently,
making a change of the coupling parameter
${c \rightarrow {c \over {S_N (K,j)}}}$, the stability multiplier
$\lambda_j$ for the nonglobal-coupling case of range $K$  becomes the
same as that for the global-coupling case.

For each mode with nonzero index $j$, we consider a region in the
parameter plane, in which a synchronous orbit is stable against the
perturbations of both modes with indices $0$ and $j$. This stable
region is bounded by four bifurcation curves determined by the
equations $\lambda_0 = \pm 1$ and $\lambda_j= \pm 1$, and it will be
denoted by $U_N$. For the case of global coupling, those stable
regions coincide, irrespectively of $N$ and $j$, because all the
asynchronous stability multipliers $\lambda_j$'s $(j \neq 0)$ are the
same, independently of $N$. The stable region for this global-coupling
case will be denoted by $U_G$. Note that $U_G$ itself is just the
stability region of the synchronous orbit, irrespectively of $N$,
because the synchronous orbit is stable against the perturbations of
all the synchronous and asynchronous modes in the region $U_G$.

However, the stable regions $U_N$'s vary depending on the coupling
range $K$ and the mode number $j$ for the nonglobal-coupling cases,
i.e., $U_N=U_N(K,j)$. To find the stability region of a synchronous
orbit in an $N$-coupled map with a given $K$, one may start with the
stability region $U_G$ for the global-coupling case. Rescaling the
coupling parameter $c$ by a scaling factor $1 \over S_N(K,j)$ for
each nonzero $j$, the stable region $U_G$ is transformed into a
stable region $U_N(K,j)$. Then the stability region of the
synchronous orbit is given by the intersection of all such stable
regions $U_N$'s. A significant change in the stability diagram
of the synchronous orbits of period $p^n$ $(n=0,1,2,\dots)$ may or
may not occur according as the coupling is linear or not, as will
be seen below.

As the first example, we study the linearly-coupled, nearest-neighbor
coupling case with $K=1$, in which the coupling function is
\begin{equation}
g(x_1,\dots,x_N) = {c \over 3} (x_2 + x_N - 2 x_1)\;{\rm for}\;
N \geq 3.
\end{equation}
This kind of coupling can be regarded as a generalized version of
Eq.~(\ref{eq:TLC}) for the linearly-coupled case for $N=2$.
(Note that as mentioned above, the cases of $N=2$ and $3$ correspond
to the global-coupling case.)
For $K=1$, the scaling factor $1 \over S_N(K,j)$ of Eq.~(\ref{eq:SF})
becomes
\begin{equation}
S_N(1,j) = {4 \over 3} \sin^2({{\pi j} \over N}).
\end{equation}
This scaling factor $1 \over S_N(1,j)$ has its minimum value,
${3 \over 4} ({3 \over {4 \cos^2{\pi \over {2N}}}})$
at $j_{\rm min} = { N \over 2} ({{N-1} \over 2})$ for even (odd) $N$.
We also note that the minimum value for odd $N$ depends on $N$, but
as $N \rightarrow \infty$ it converges to the minimum value for the
case of even $N$.

Rescaling the coupling parameter $c$ with the scaling factor $1 \over
S_N(1,j)$, the stability region $U_G$ for the $N=2$ and $3$ cases of
global coupling is transformed into the region $U_N(1,j)$ for $N>3$.
Then, as a result of the intersection of all such regions
$U_N(1,j)$'s, only the region $U_N(1,j_{\rm min})$ including a $c=0$
line segment is left as the stability region of a synchronous orbit.
Consequently, of the infinite number of period $p$-tupling
routes for the global-coupling case, only the $Z_p$ route ending at
the zero-coupling critical point $(A_{\infty}^{(p)},0)$ remains. Thus
only the zero-coupling point is left as a critical point in the
parameter plane. An example for the period-tripling case with $p=3$
is shown in Fig.~\ref{MLCSD}. The largest stability region
corresponds to the stability region for the $N=2$ and $3$ cases of
the global coupling, while the smallest one corresponds to that
for the case of even $N$ $(N>2)$. Between them, there exist
stability regions for the case of odd $N$ $(N>3)$ (e.g., see the
$N=5$ case in Fig.~\ref{MLCSD}).

We now examine the CB near the zero-coupling critical point for the
case of the linearly-coupled, nearest-neighbor coupling.
Consider a self-similar sequence of parameters $(A_n,c_n)$, at which
the synchronous orbit  of level $n$ has some  given stability
multipliers, in the $Z_p$ route for the global-coupling case.
Rescaling the coupling parameter with the factor
$1 \over S_N(1,j_{\rm min})$, this sequence is transformed into a
self-similar one for the case of the linearly-coupled,
nearest-neighbor coupling. Thus the ``width''  of each stability
region in the  $Z_p$ route for the case of the global coupling is
reduced to that for the case of the linearly-coupled, nearest-neighbor
coupling by the scaling factor $1 \over S_N(1,j_{\rm min})$, while
the ``heights'' of all stable regions in the $Z_p$ route remain
unchanged \cite{note2}. It is therefore obvious that the critical
scaling behavior near the zero-coupling critical point for the case
of the linearly-coupled, nearest-neighbor coupling is the same as that
for the global coupling case.
That is, the height and width $h_n$  and $w_n$ of the stability region
of level $n$ geometrically contract in the limit of large $n$,
\begin{equation}
h_n  \sim  \delta^{-n},\;\;w_n \sim \alpha^{-n}\;\;{\rm
for\;large\;} n,
\end{equation}
where $\delta$ and $\alpha$ are the scaling factors of the
nonlinearity and coupling parameters, respectively.
As an example, see again Fig.~\ref{MLCSD} and
note that Figs.~\ref{MLCSD}(a), \ref{MLCSD}(b) and \ref{MLCSD}(c)
nearly coincide near the zero-coupling critical point except for
small numerical differences.

The results of the linearly-coupled, nearest-neighbor coupling
with $K=1$ extend to all the other linearly-coupled,
nonglobal-coupling cases with $1 < K < {N \over 2} ({{N-1} \over 2})$
for even (odd) $N$. For each nonglobal-coupling case with $K>1$, we
first consider a mode with index $j_{\rm min}$ for which the scaling
factor $1 \over S_N(K,j)$ becomes the smallest one.
Here the value of $j_{\rm min}$ varies depending on the range $K$.
Like the $K=1$ case, only the region $U_N(K,j_{\rm min})$ including a
$c=0$ line segemnt is left as the stability region of a synchronous
orbit. Consequently, only the zero-coupling point remains as a
critical point in the parameter plane, and the CB near the
zero-coupling critical point is also the same as that for the
global-coupling case.

Finally, as an example of the nonlinear coupling, we study the
diffusively-coupled, nearest-neighbor coupling case with $K=1$,
in which the coupling function is given by
\begin{equation}
g(x_1,\dots,x_N) = {c \over 3}
[f(x_2) + f(x_N) - 2 f(x_1)]\;{\rm for}\;
N \geq 3.
\end{equation}
This kind of coupling can be regarded as a generalized version of
Eq.~(\ref{eq:DCFTN}) for the dissipatively-coupled case for $N=2$.
For this diffusively-coupled case, the stability multipliers
of Eq.~(\ref{eq:NGCASSM}) for a synchronous $q$-periodic orbit
of level $n$ ($q=p^n$)
become
\begin{equation}
\lambda_j = [1 - c S_N(1,j)]^q \lambda_0.
\end{equation}

Like the linearly-coupled case, rescaling the coupling parameter $c$
with the minimum scaling factor $1 \over S_N(1,j_{\rm min})$, each
stability region $U_G$ of level $n$ for the global-coupling case is
transformed into the stability region
[i.e., the region $U_N(1,j_{\rm min})]$ of level $n$
for the diffusively-coupled case. The stability region
$U_N(1,j_{\rm min})$ is bounded by four bifurcation curves
determined by the equations $\lambda_j = \pm 1$ for $j=0,j_{\rm min}$.
An infinite sequence of such stability regions converges to a critical
line segment joining two points $c^*_l$ $(=0)$and $c^*_r$
$(={2 \over {S_N(1,j_{\rm min})}})$ on the $A=A_\infty^{(p)}$ line.
Consequently, one critical line segment constitutes the critical set,
like the global-coupling case. An example for the period-tripling case
for even $N$ $(N>2)$ is shown in Fig.~\ref{MDCSD}.

As shown above, the stability diagram for the diffusively-coupled case
is essentially the same as that for the globally-coupled case, except
for the scale in the coupling parameter $c$. Hence the critical
scaling behavior of the nonlinearity and coupling parameters and so on
becomes the same as that for the global-coupling case (for details on
the CB, refer to Sec.~\ref{sub:DCM}). Finally, we briefly discuss the
critical asynchronous stability multipliers. At the zero-coupling
critical point $(A^{(p)}_\infty,0)$ and interior critical points, they
are the same as those for the global-coupling case (i.e., $\lambda_1^*
= \cdots = \lambda_{N-1}^* = \lambda_0^*$ at $(A^{(p)}_\infty,0)$, and
$\lambda_1^* = \cdots = \lambda_{N-1}^* = 0$ at interior critical
points.). However, at the right end $(A^{(p)}_\infty,c^*_r)$,
the critical asynchronous stability multiplier for $j=j_{\rm min}$
is given by $\lambda^*_{j_{\rm min}} = \lambda^*_0$ (even $q$)
and $- \lambda^*_0$ (odd $q$), but all other $\lambda^*_j$'s
$(j \neq j_{\rm min})$ are zero, which is somewhat different
from that for the global-coupling case [i.e., $\lambda^*_1=
\cdots=\lambda^*_{N-1}=\lambda^*_0$ (even $q$) and $- \lambda^*_0$
(odd $q$)]. Like the linearly-coupled case, the results for the
diffusively-coupled, nearest-neighbor coupling case
with $K=1$ can be also extended to all the other
diffusively-coupled, nonglobal-coupling cases with
$1 < K < {N \over 2} ({{N-1} \over 2})$ for even (odd) $N$.

\section{summary}
\label{sec:SUM}

The CB of all period $p$-tuplings $(p=2,3,4,\dots)$ are studied in
$N$ $(N=2,3,4,\dots)$ symmetrically coupled 1D maps.
The two-coupled case with $N=2$ is first investigated by a
renormalization method. We find three (five) kinds of fixed points
of the renormalization operator and their relevant CE's
associated with coupling perturbations for the case of even (odd) $p$.
We next consider two kinds of couplings, linear and nonlinear
couplings. As examples of the linear- and nonlinear-coupling cases,
we study the linearly- and dissipatively-coupled maps, respectively,
and confirm the renormalization results. The structure of the
critical set varies depending on the nature of coupling. For the
linearly-coupled case, an infinite number of the critical line
segments, together with the zero-coupling critical point, constitute
the critical set, while for the dissipatively-coupled case, the
critical set consists of the only one critical line segemnt. The CB
also depends on the position on the critical set. For even (odd)
$p$, three (four) kinds of fixed points govern the CB for the
linearly-coupled case, while only two (three) kinds of fixed points
govern the CB for the dissipatively-coupled case.
Finally, the results of the two coupled maps are extended to many
coupled maps with $N \geq 3$, in which the CB depends on the range of
coupling. In the global-coupling case, both the structure of the
critical set and the CB are the same as those of the two-coupled case,
independently of $N$. However, for the nonglobal-coupling case, a
significant change in the structure of the critical set may or may
not occur according as the coupling is linear or not.
For the linearly-coupled case, of the infinite
number of period $p$-tupling routes for the global-coupling case,
only the route ending at the zero-coupling critical point is left in
the parameter plane. On the other hand, for the diffusively-coupled
case, one critical line segment constitutes the critical set, as in
the globally-coupled case.

\acknowledgments
This work was supported by the Basic Science Research Institute
Program, Ministry of Education, Korea, Project No. BSRI-95-2401. The
author thanks Professor R. Fox at the Georgia institute of Technology
for his hospitality.

%
%  End of Reference
%

\begin{table}
\caption{ Reduced coupling fixed functions $G^*(x)$, relevant CE's
$\nu$ and the critical asynchronous stability multipliers
$\lambda_1^*$ in all the period $p$-tupling cases are shown for the
case of two coupled maps. In the
second column, ``e'' (``o'') denotes the even (odd) period-$p$
tuplings. The first three ones for $G^*(x)$ exist for all $p$, while
the last two ones exist only for odd $p$. Note also that the case
$G^*(x) = {1 \over 2} {f^*}'(x)$ has no relevant CE's,
and $\alpha$ and $\lambda^*$ are the orbital scaling factor and the
critical stability multiplier for the 1D case, respectively.
        }
\label{table1}
\begin{tabular}{cccc}
$G^*(x)$ & $p$ & $\nu$ & $\lambda_1^*$ \\
\tableline
0 & e, o &$\alpha$, $p$   &  $\lambda^*$  \\
${1 \over 2}$ ${f^*}'$ & e, o & nonexistent &  0 \\
${1 \over 2} [{f^*}'(x)-1]$ &e, o & $p$ &  1 \\
${1 \over 2} [{f^*}'(x) +1]$ &o &  $p$ &  -1 \\
${f^*}'(x)$ &o & $\alpha$, $p$ &  $-\lambda^*$
\end{tabular}
\end{table}

\begin{table}
\caption{ In the $Z_3$ route, we followed a sequence of parameters
          $(A_n,c_n)$ at which the pair of stability multipliers
          $(\lambda_{0,n},\lambda_{1,n})$
          of the orbit of period $3^n$ is $(-1,1)$. This
          sequence converges to the zero-coupling critical point
          $(A_\infty^{(3)},0)$ with the scaling factors shown in the
          second and third columns.
        }
\label{table2}
\begin{tabular}{ccc}
$n$ & $\delta_n$ & $\mu_n$ \\
\tableline
3 & 55.264\,789\,71  &  -9.272\,61 \\
4 & 55.245\,771\,51  &  -9.279\,20 \\
5 & 55.247\,110\,93 &   -9.276\,71 \\
6 & 55.247\,020\,84 &   -9.277\,55 \\
7 & 55.247\,026\,98 &    -9.277\,27 \\
8 & 55.247\,026\,56 &   -9.277\,36 \\
9 & 55.247\,026\,59 &   -9.277\,33
\end{tabular}
\end{table}

\begin{table}
\caption{For the case of the $Z_3$ route, scaling factors $\mu_{1,n}$
         and $\mu_{2,n}$ in the two-term scaling for the coupling
         parameter are shown in the second and third columns,
         respectively. A product of them, ${\mu^2_{1,n}} \over
         {\mu_{2,n}}$, is shown in the fourth column.
        }
\label{table3}
\begin{tabular}{cccc}
$n$ & $\mu_{1,n}$ & $\mu_{2,n}$ & ${\mu^2_{1,n}} \over {\mu_{2,n}}$ \\
\tableline
4 & -9.277\,396\,06  &  24.578\,89 &  3.501\,79 \\
5 &  -9.277\,337\,31 &  27.768\,52 &  3.099\,52 \\
6 & -9.277\,341\,36 &   28.502\,38 &  3.019\,72 \\
7 & -9.277\,341\,10 &   28.650\,77 &  3.004\,08 \\
8 & -9.277\,341\,12 &   28.681\,53 &  3.000\,85
\end{tabular}
\end{table}

\begin{table}
\caption{In the period-tripling case, scaling factors $r_{1,n}$ and
         $r_{2,n}$ in the two-term scaling for the slope $S_n$ of the
         asynchronous stability multiplier $\lambda_{1,n}$ at the
         zero-coupling critical point are shown in the second and
         third columns, respectively.
        }
\label{table4}
\begin{tabular}{ccc}
$n$ & $r_{1,n}$ & $r_{2,n}$  \\
\tableline
4 & -9.277\,335\,543\,4  &  2.927\,8  \\
5 &  -9.277\,341\,501\,0 &  2.984\,5  \\
6 & -9.277\,341\,089\,3 &   2.996\,7  \\
7 & -9.277\,341\,117\,4 &   2.999\,3  \\
8 & -9.277\,341\,116\,8 &   2.999\,5
\end{tabular}
\end{table}

\begin{table}
\caption{We followed, in the leftmost $L_3$ route, two self-similar
         sequences of parameters. One sequence of parameters
         $(A_n,c_n)$, at which $\lambda_{0,n}=-1$ and
         $\lambda_{1,n}=0.8$, converges to the left end
         $(A_\infty^{(3)},c_l)$ of the critical line segment with the
         scaling factor $\mu_n$ of the coupling parameter shown in
         the second column. The other sequence of parameters
         $(A_n,c_n)$, at which $\lambda_{0,n}=-1$ and $\lambda_{1,n}
         = - 0.8$, converges to the right end $(A_\infty^{(3)},c_r)$
         with the scaling factor $\mu_n$ of the coupling parameter
         shown in the fourth column. In both cases the scaling factors
         are the same. The convergence of the sequence
         $\{ \lambda_{1,n} \}$ to its limit values $\lambda_1^*$ at
         the left (right) end is also shown in the third (fifth)
         column. Note that the
         values of $\lambda_1^*$'s at both ends are different.
        }
\label{table5}
\begin{tabular}{ccccc}
$n$ & $\mu_n$ & $\lambda_{1,n}$ & $\mu_n$ & $\lambda_{1,n}$  \\
\tableline
3 & 3.664\,5 & 0.996\,763\,57 & 1.836\,2  & -1.010\,285\,72 \\
4 & 2.946\,7  & 1.000\,444\,09 & 3.057\,3 & -0.998\,599\,423 \\
5 & 2.990\,8 & 0.999\,939\,19 & 2.968\,2 &  -1.000\,191\,98  \\
6 & 2.994\,1 & 1.000\,000\,33 & 2.997\,0 &  -0.999\,973\,71  \\
7 & 2.998\,3 & 0.999\,998\,86 & 2.998\,0 &  -1.000\,003\,60  \\
8 & 2.999\,4 & 1.000\,000\,16 & 2.999\,5 &  -0.999\,999\,51 \\
9 & 2.999\,8 & 0.999\,999\,98 & 2.999\,8 &  -1.000\,000\,07
\end{tabular}
\end{table}

\begin{table}
\caption{ In the $Z_4$ route, we followed a sequence of parameters
          $(A_n,c_n)$ at which the pair of stability multipliers
          $(\lambda_{0,n},\lambda_{1,n})$
          of the orbit of period $4^n$ is $(-1,1)$. This
          sequence converges to the zero-coupling critical point
          $(A_\infty^{(4)},0)$ with the scaling factors shown in
          the second and third columns.
        }
\label{table6}
\begin{tabular}{ccc}
$n$ & $\delta_n$ & $\mu_n$ \\
\tableline
2 & 981.297\,281  &  -38.841\,91 \\
3 & 981.596\,790  &  -38.816\,17 \\
4 & 981.594\,965  &  -38.819\,38 \\
5 & 981.594\,977  &  -38.819\,04 \\
6 & 981.594\,976  &  -38.819\,08
\end{tabular}
\end{table}

\begin{table}
\caption{For the case of the $Z_4$ route, scaling factors $\mu_{1,n}$
         and $\mu_{2,n}$ in the two-term scaling for the coupling
         parameter are shown in the second and third columns,
         respectively. A product of them, ${\mu^2_{1,n}}
         \over {\mu_{2,n}}$, is shown in the fourth column.
        }
\label{table7}
\begin{tabular}{cccc}
$n$ & $\mu_{1,n}$ & $\mu_{2,n}$ & ${\mu^2_{1,n}} \over {\mu_{2,n}}$ \\
\tableline
3 & -38.819\,021\,99  &  311.92 &  4.83 \\
4 & -38.819\,074\,56 &   372.98 &  4.04 \\
5 & -38.819\,074\,24 &   377.22 &  3.99
\end{tabular}
\end{table}

\begin{table}
\caption{In the period-quadrupling case, scaling factors $r_{1,n}$ and
         $r_{2,n}$ in the two-term scaling for the slope $S_n$ of the
         asynchronous stability multiplier $\lambda_{1,n}$ at the
         zero-coupling critical point are shown in the second and
         third columns, respectively.
        }
\label{table8}
\begin{tabular}{ccc}
$n$ & $r_{1,n}$ & $r_{2,n}$  \\
\tableline
2 & -38.819\,023\,65  &  3.593  \\
3 & -38.819\,074\,65 &  3.972  \\
4 & -38.819\,074\,29 &  3.998
\end{tabular}
\end{table}

\begin{table}
\caption{We followed, in the leftmost $L_4$ route, two self-similar
         sequences of parameters $(A_n,c_n)$, at which
         $\lambda_{0,n}=-1$ and $\lambda_{1,n}=0.8$. One sequence
         converges to the left end $(A_\infty^{(4)},c_l)$ of the
         critical line segment with the scaling factor $\mu_n$ of the
         coupling parameter shown in the second column. The other
         sequence converges to the right end $(A_\infty^{(4)},c_r)$
         with the scaling factor $\mu_n$ of the coupling parameter
         shown in the fourth column. In both cases the scaling factors
         are the same. The convergence of the sequence
         $\{ \lambda_{1,n} \}$ to its limit values
         $\lambda_1^*$ at the left (right) end is also shown in the
         third (fifth) column. Note that the
         values of $\lambda_1^*$'s at both ends are the same.
        }
\label{table9}
\begin{tabular}{ccccc}
$n$ & $\mu_n$ & $\lambda_{1,n}$ & $\mu_n$ & $\lambda_{1,n}$  \\
\tableline
2 & 5.013 & 1.010\,032\,65 & 50.70 & 0.987\,196\,90 \\
3 & 4.051 & 0.999\,630\,89 & 3.674 &  1.000\,476\,66\\
4 & 3.963 & 1.000\,013\,65 & 3.978 &  0.999\,982\,38  \\
5 & 3.992 & 0.999\,999\,50 & 3.993 &  1.000\,000\,65  \\
6 & 3.998 & 1.000\,000\,02 & 3.998 &  0.999\,999\,98
\end{tabular}
\end{table}

\begin{table}
\caption{Independent reduced coupling fixed functions $G^*(x)$,
relevant CE's $\nu$ and independent critical asynchronous stability
multipliers
$\lambda_1^*$ in all the period $p$-tupling cases are shown for the
case of $N$ globally coupled maps. The first three ones for $G^*(x)$
exist for all $p$, whereas the last two ones exist only for odd $p$.
Here $\alpha$ and $\lambda^*$ are the orbital scaling factor and the
critical stability multiplier for the 1D case, respectively. Note
that $\nu$ and $\lambda^*_1$ for each $G^*(x)$ are also the same as
those for the two-coupled case, independently of $N$.
        }
\label{table10}
\begin{tabular}{ccc}
$G^*(x)$ & $\nu$ & $\lambda_1^*$ \\
\tableline
0 & $\alpha$, $p$   &  $\lambda^*$  \\
${1 \over N}$ ${f^*}'$  & nonexistent &  0 \\
${1 \over N} [{f^*}'(x)-1]$ & $p$ &  1 \\
${1 \over N} [{f^*}'(x) +1]$ &  $p$ &  -1 \\
${2 \over N} {f^*}'(x)$& $\alpha$, $p$ &  $-\lambda^*$
\end{tabular}
\end{table}

\begin{figure}
\caption{Stability diagram of the synchronous $3^n$-periodic orbits
         of the lowest two levels $n=0,1$ in two linearly coupled
         maps. Each periodic orbit of level $n$ is born via its own
         saddle-node bifurcation. Its stable regions, denoted by
         $q=3^n$, are bounded by four bifurcation curves
         determined by $\lambda_i = \pm 1$ for $i=0,1$.
         The horizontal and non-horizontal solid (short-dashed)
         boundary lines (i.e., the $\lambda_0=-1$ $(1)$ and
         $\lambda_1=-1$ $(1)$ bifurcation curves) correspond to the
         synchronous and asynchronous period-doubling (synchronous
         saddle-node and asynchronous pitchfork) bifurcation curves,
         respectively.
     }
\label{SDT1}
\end{figure}

\begin{figure}
\caption{Stability diagram of the synchronous $3^n$-periodic
         $(n=1,2,3)$ orbits of level $n$ in two linearly coupled maps.
         Each periodic orbit of level $n$ is born via its own
         saddle-node bifurcation. Its stable regions, denoted by
         $q=3^n$, are bounded by four bifurcation curves
         determined by $\lambda_i = \pm 1$ for $i=0,1$.
         The solid and short-dashed boundary lines represent the
         same as those in Fig.~1. The stability diagrams starting from
         the central, right, and left stability regions of level $1$
         are shown in (a), (b), and (c), respectively. See the text for
         other details.
     }
\label{SDT2}
\end{figure}

\begin{figure}
\caption{Stability regions of the synchronous orbits of period
         $q=3^n$ $(n=1,2,3)$ in two linearly coupled maps.
         The cases $n=1,2,$ and $3$ are shown in (a), (b), and (c),
         respectively. Its stable regions, denoted by
         $q=3^n$, are bounded by four bifurcation curves
         determined by $\lambda_i = \pm 1$ for $i=0,1$.
         The solid and short-dashed boundary lines represent the
         same as those in Fig.~1. The scaling factors used in (b) and
         (c) are $\delta = 55.247\,026$ and $\alpha = -9.277\,341$.
     }
\label{PS}
\end{figure}

\begin{figure}
\caption{Plots of the asynchronous stability multipliers
         $\lambda_{1,n}(A^{(3)}_\infty,c)$ versus $c$
         near the zero-coupling critical point for the
         synchronous orbits of period $q=3^n$ ($n=2,3,4$).
     }
\label{ASM1}
\end{figure}

\begin{figure}
\caption{Plots of the asynchronous stability multipliers
         $\lambda_{1,n}(A^{(3)}_\infty,c)$ versus $c$
         near the leftmost critical line for the synchronous
         orbits of period $q=3^n$ $(n=2,3,4)$.
     }
\label{ASM2}
\end{figure}

\begin{figure}
\caption{Plot of the asynchronous Lyapunov exponent
         $\sigma_1(A^{(3)}_\infty,c)$ versus $c$ near the leftmost
         critical line. The plot consists of $200$ $c$ values, each of
         which is obtained by iterating the map $100\,000$ times to
         eliminate transients and then averaging over another
         $500\,000$ iterations.
     }
\label{ALexp1}
\end{figure}

\begin{figure}
\caption{Attractors near the leftmost critical line: (a) a synchronous
         attractor inside the critical line and asynchronous
         attractors outside the critical line for (b) $c=-3.590\,5$
         and (c) $c=-3.482\,4$. For each case, the map is iterated
         $100\,000$ times to eliminate transients and the next
         $10\,000$ iterations are plotted.
     }
\label{Desyn}
\end{figure}

\begin{figure}
\caption{Plots of the asynchronous Lyapunov exponent
         $\sigma_1(A^{(3)}_\infty,c)$ versus $c$ near the (a) left
         and (b) right ends of the leftmost critical line.
         Each plot consists of $50$ $\varepsilon$ values, each of
         which is obtained by iterating the map $100\,000$ times
         to eliminate transients
         and then averaging over another $500\,000$ iterations.
     }
\label{ALexp2}
\end{figure}

\begin{figure}
\caption{Stability diagram of the synchronous $4^n$-periodic orbits
         of the lowest two levels $n=0,1$ in two linearly coupled
         maps. Each periodic orbit of level $n$ is born via its own
         saddle-node bifurcation. Its stable regions, denoted by
         $q=4^n$, are bounded by four bifurcation curves
         determined by $\lambda_i = \pm 1$ for $i=0,1$.
         The solid and short-dashed boundary lines represent the
         same as those in Fig.~1.
     }
\label{SDQ1}
\end{figure}

\begin{figure}
\caption{Stability diagram of the synchronous $4^n$-periodic
         orbits of level $n$ $(n=1,2,3)$ in two linearly coupled maps.
         Each periodic orbit of level $n$ is born via its own
         saddle-node bifurcation. Its stable regions, denoted by
         $q=4^n$, are bounded by four bifurcation curves
         determined by $\lambda_i = \pm 1$ for $i=0,1$.
         The solid and short-dashed boundary lines represent the
         same as those in Fig.~1. The stability diagrams starting from
         the central, right, and left stability regions of level $1$
         are shown in (a), (b), and (c), respectively. See the text for
         other details.
     }
\label{SDQ2}
\end{figure}

\begin{figure}
\caption{Stability regions of the synchronous orbits of period
         $q=4^n$ $(n=1,2,3)$ in two linearly coupled maps.
         The cases $n=1,2,$ and $3$ are shown in (a), (b), and (c),
         respectively. Its stable regions, denoted by
         $q=4^n$, are bounded by four bifurcation curves
         determined by $\lambda_i = \pm 1$ for $i=0,1$.
         The solid and short-dashed boundary lines represent the
         same as those in Fig.~1. The scaling factors used in (b) and
         (c) are $\delta = 981.594\,976$ and $\alpha = -38.819\,074$.
     }
\label{QPS}
\end{figure}

\begin{figure}
\caption{Plots of the asynchronous stability multipliers
         $\lambda_{1,n}(A^{(4)}_\infty,c)$ versus $c$
         near the zero-coupling critical point for the
         synchronous orbits of period $4^n$ ($n=1,2,3$).
     }
\label{QASM1}
\end{figure}

\begin{figure}
\caption{Plots of the asynchronous stability multipliers
         $\lambda_{1,n}(A^{(4)}_\infty,c)$ versus $c$
         near the leftmost critical line for the synchronous
         orbits of period $4^n$ ($n=2,3,4$).
     }
\label{QASM2}
\end{figure}

\begin{figure}
\caption{Plot of the asynchronous Lyapunov exponent
         $\sigma_1(A^{(4)}_\infty,c)$ versus $c$ near the leftmost
         critical line. The plot consists of $200$ $c$ values, each of
         which is obtained by iterating the map $100\,000$ times
         to eliminate transients and then averaging over another
         $500\,000$ iterations.
     }
\label{QALexp1}
\end{figure}

\begin{figure}
\caption{Attractors near the leftmost critical line: (a) a synchronous
         attractor inside the critical line and asynchronous
         attractors outside the critical line for (b) $c=-3.877\,059$
         and (c) $c=-3.888\,059$. For each case, the map is iterated
         $100\,000$ times to eliminate transients and the next
         $10\,000$ iterations are plotted.
     }
\label{QDesyn}
\end{figure}

\begin{figure}
\caption{Plots of the asynchronous Lyapunov exponent
         $\sigma_1(A^{(4)}_\infty,c)$ versus $c$ near the (a) left
         and (b) right ends of the leftmost critical line.
         Each plot consists of $50$ $\varepsilon$ values, each of
         which is obtained by iterating the map $100\,000$ times
         to eliminate transients
         and then averaging over another $500\,000$ iterations.
     }
\label{QALexp2}
\end{figure}

\begin{figure}
\caption{Stability diagram of the synchronous orbits for the
         (a) period-tripling ($p=3$) and (b) period-quadrupling
         $(p=4)$ cases in two dissipatively coupled maps. Each
         periodic orbit of level $n$ is born via its own saddle-node
         bifurcation. Its stable regions, denoted by $q=p^n$
         $(n=1,2,3)$, are bounded by four bifurcation curves
         determined by $\lambda_i = \pm 1$ for $i=0,1$.
         The solid and short-dashed boundary lines represent the
         same as those in Fig.~1.
     }
\label{DCSD}
\end{figure}

\begin{figure}
\caption{Fixed points of the recurrence equations (a) (3.35)
         and (b) (3.36). The intersection points between the curves
         of Eqs.~(3.35) and (3.36) and the line $c_n = c_{n-1}$ are
         just the fixed points, denoted by solid circles.
     }
\label{DCFP}
\end{figure}

\begin{figure}
\caption{Plots of the asynchronous stability multipliers
         $\lambda_{1,n}(A_\infty^{(p)},c)$ versus $c$ for the
         synchronous orbits of period $p^n$ $(n=1,2,3)$
         near the critical line for the (a) period-tripling
         $(p=3)$ and (b) period-quadrupling $(p=4)$ cases in two
         dissipatively coupled maps.
     }
\label{DCSM}
\end{figure}

\begin{figure}
\caption{Asynchronous attractors outside the critical line for
         (a) $c=-0.001\,5$ in the period-tripling case and (b)
         $c=-0.000\,1$ for the period-quadrupling case. For each case,
         the map is iterated $100\,000$ times to eliminate transients
         and the next $10\,000$ iterations are plotted.
         }
\label{DCDesyn}
\end{figure}

\begin{figure}
\caption{Stable regions of the synchronous $3^n$-periodic orbits of
         level $n=1,2,3$ in $N$ $(N=2,3,4,\dots)$ coupled 1D maps for
         the linearly-coupled, nearest-neighbor coupling case with
         $K=1$. The cases of $n=1$, $2$, and $3$ are shown in (a), (b),
         and (c), respectively. The symbols $q_j^{PD}$, $q_j^{SN}$, and
         $q_j^{PF}$ with $j=0$ and $j_{\rm min}$ represent the
         period-doubling, saddle-node, and pitchfork bifurcation lines
         associated with the stability of a synchronous orbit of period
         $q$ against the perturbation of the $j$th mode.
         The scaling factors used in (b) and (c) are
         $\delta = 55.247\,026$ and $\alpha = -9.277\,341$.
         For other details, see the text.
         }
\label{MLCSD}
\end{figure}

\begin{figure}
\caption{Stable regions of the synchronous $3^n$-periodic orbits of
         level $n=1,2,3$ in even-$N$ $(N>2)$ coupled 1D maps for
         the diffusively-coupled, nearest-neighbor coupling case with
         $K=1$. The symbols $q_j^{PD}$, $q_j^{SN}$, and
         $q_j^{PF}$ with $j=0$ and $j_{N \over 2}$ represent the
         period-doubling, saddle-node, and pitchfork bifurcation lines
         associated with the stability of a synchronous orbit of period
         $q$ against the perturbation of the $j$th mode.
         }
\label{MDCSD}
\end{figure}


\begin{references}
\bibitem{note} In deriving Eq.~(\ref{eq:RRE}), we use the relation,
$F_{n,1}^{(p)}(x) + F_{n,2}^{(p)}(x) = f_n^{(p)'}(x)$, where
$F^{(p)}_{n,1}(x) \equiv \partial F^{(p)}_n (x,y) /
\partial x|_{y=x}$ and $F^{(p)}_{n,2}(x) \equiv \partial F^{(p)}_n
(x,y) / \partial y|_{y=x}$.
\bibitem{note2} We define the ``height'' $h_n$ of the stable region of level
$n$ in the $Z_3$ or $Z_4$ route as the length of the $c=0$ line segment inside
the stable region. The ``width'' $w_n$ is also defined as the length of the
boundary line segment associated with the synchronous period-doubling
bifurcation (i.e., the curve determined by the equation $\lambda_{0,n}=-1$).
\end{references}
\end{document}